# Rule-based High-level Hardware-RTL Synthesis of Algorithms, Virtualizing Machines, and Communication Protocols with FPGAs based on Concurrent Communicating Sequential Processes and the ConPro Synthesis Framework

Stefan Bosse[1,2.3]
[1]University of Bremen, Dept. of Mathematics and Computer Science, Bremen, Germany
[2]University of Siegen, Dept. of Mechanical Engineering, Siegen, Germany
[3]Institute for Digitization, Bremen, Germany

*sbosse@uni-bremen.de*

*Abstract* — Virtualization is the abstraction of details. Algorithms and programming languages provide abstraction, too. Virtualization of hardware and embedded systems is becoming more and more important in heterogeneous environments and networks, e.g., distributed and material-integrated sensor networks. Communication and data processing with a broad range of hardware and low-level protocols can be unified and accessed uniquely by introducing virtualization layers implemented directly in hardware on chip. Hardware design is today still component-driven (like a circuit board), rather than transforming algorithms as an abstraction layer directly into hardware designs. Programs and protocols are algorithms, so do not handle them as devices like in traditional high-level synthesis design flows! Complex reactive systems with dominant and complex control paths play an increasing role in SoC-design. The major contribution to concurrency appears at the control path level. This article gives an in-depth introduction to SoC-design methodology using the Highest-Level Synthesis ConPro compiler framework and a process-oriented programming language that provides a programming model based on concurrently executing and communicating sequential processes (CCSP) with an extensive set of interprocess-communication primitives. Circuits are modelled and programmed on an algorithmic level, more convenient and natural than component-driven designs. Extended case studies of a smart communication protocol router and an advanced stack-based processor providing a programmatical virtualization layer are shown and evaluated. Both are used together as a smart node architecture deployed in high density sensor-actuator-networks, e.g., for material-integrated intelligent systems.

*Software and Code* — All software and code examples (case studies) can be downloaded from:
`https://github.com/bsLab/conpro2`

## Table of Content

# I Introduction

The development of low-power microprocessors and microcontrollers deployed in embedded systems and sensor nodes has shown significant improvement in recent years with respect to the ratio of energy power to computational power. But, there is still an increasing requirement for the deployment of System-On-Chip-Designs (SoC) using Application-Specific Digital Circuits, with increasing complexity and serving low-power, real-time, and miniaturization constraints. The structural decomposition of such a SoC into independent submodules requires smart communication networks and protocols (Network-on-Chip, NoC) that serve chip area and power limitations. Traditionally, SoCs are composed of microprocessor cores, memory, and peripheral components.

The Register-Transfer-Logic (RTL) on architecture and hardware level must be derived from the algorithmic level, requiring a raise of abstraction of RTL **[ZHU01]**. But in general, massive parallel systems require modelling of parallelism and concurrency both on the control- and data-path levels. Digital logic systems and Field Programmable Gate Arrays (FPGA) as a rapid prototyping tool are preferred for the exploration and implementation of parallelism on the control- and data-path levels.

The hardware design and synthesis of program-controlled microprocessors and microcontrollers with a specific Instruction Set Architecture (ISA) is widely used. The ISA is typically used by software compilers. Virtual Machines (VM), in contrast, define their own ISA independent from the underlying processor ISA (regarding system and process virtual machines, e.g., qemu). VM are commonly software. But virtualization as a method for abstraction, simplification, and unification, can also implemented directly in hardware, creating Virtualizing Machines (ViM), addressed in this work. A ViM is commonly characterised by the capability to update and extend the ISA seamlessly. Algorithms, in general, can be considered as virtualization, too, since they implement some kind of function that maps input data on output data.

There are plenty of synthesis tools and modelling languages available to create SoC circuits, but they must be driven primarily by System-C models and programming languages **[MEE12]**. There are software and hardware designers. Software designers have to use existing hardware under specific data type, data range, and data flow, instruction set, and control flow constraints given by the hardware. The hardware designer has full freedom to determine ranges, data types, data flows, operations, and control path design, including exploitation of parallelism. Common hardware modelling languages like Verilog or VHDL give designers this freedom, but with the significant disadvantage of requiring expert knowledge and exploding model complexity. A good High-level synthesis (HLS) tool should allow the designer to make free choices of data- and control path design and let them validate these choices at the source code level.

HLS's capabilities for design exploration are one of its key advantages. HLS enables the examination of many architectures to select the optimal one in light of the design requirements, rather than making architectural selections in advance and manually writing the design of the Register-Transfer Level (RTL). The way that HLS tools direct this exploring process varies greatly. Maximum separation between the source code (behavior) and design restrictions (architecture) is desirable for clarity and portability. Another important feature that a HLS tool should provide is scalability and adaptability of designs and architectures.**[MAN17]**

Besides designing complete and self-contained application- or domain-specific circuits, accelerators are prominent examples of SoCs extending existing software-controlled systems. But the regularity and reusability of accelerators across platforms are diminished by heterogeneity and hardware specialization, which hinder accelerator design and integration **[PIC22]**. HLS enables rich and fast Design-Space Exploration (DSE), but it is mostly manual. Expert knowledge is required for automated design space exploration.

Using widely used and well-known programming languages for hardware design is an attractive way to reach a broader developer community. For example, Circe **[RAD02]** uses a variant of C and produces a structural hardware description in RTL VHDL. Using generic memory-mapped languages like C makes RTL hardware synthesis difficult because of the transparency of object references (using pointers), preventing RTL mapping. Additionally, concurrency models are missing in most software languages. Many attempts have been made to use C-like languages, but with limitations, such as prohibiting anonymous memory access with pointers, or by using a program-controlled (multi-) processor architecture with traditional hardware-software co-design, which is currently dominant in SoC-Design.But SoC-designs using generic or application-specific processor architectures complicate low-power designs, and concurrency is coarse grained.

Many of the highly parallel FPGA-based systems that Xilinx and its customers develop for applications like telecommunications and video processing call for the use of one or more additional microprocessors. But managing the interactions between the software running on the microprocessors and the interactions between the existing FPGA cores is challenging **[PER13]**.

With increasing complexity, higher abstraction levels are required, moving from the hardware to the algorithmic level. Programming languages that are naturally imperative are used to implement algorithms on program-controlled machines that process a sequential stream of data and control operations.Using this data-processing architecture, a higher-level imperative language can be simply mapped to a lower-level imperative machine language, which is a rule-based mapping, automatically performed by a software compiler.

But in circuit design, there is neither an existing architecture nor a low-level language that can be synthesized directly from a higher level one. Additionally, there is not pre-defined efficient computer architecture for parallel data processing systems.

Another example is PICO **[KAT02]**, which addresses the complete hardware design flow targeting SoC and customizable or configurable processors, enhanced with custom-designed hardware blocks (accelerators). C is used to model the RTL level. The program-controlled approach with processor blocks enables software compilation and unrestricted C (functions, pointers), but lacks support for true bit-scaled data objects. SPARK **[GUP04]**, a C-to-VHDL high-level framework, currently has the restrictions of no pointers, no function recursion, and no irregular control-flow jumps. It is embedded in a traditional hardware-software-co-design flow. It is based on speculative code motions and loop transformations used for exploration of concurrency. SPARK generates pure RTL. Only a single-threaded control-flow is provided. Recent software frameworks addressing Hardware-Software Co-designs include, for example, Vivaldi **[COR16]**.

Although, SystemC provides many features suitable for higher-level synthesis, it is primarily used for simulation and verification, and only a subset can be synthesized into circuits. True bit-scaled data types are supported. Threaded processes, such as those used in Forthes commercial synthesis tool, Cynthesizer **[COU08][MER08][PHI08]**, can be used to model concurrency. The transaction level (TL) is used to model inter-process communication.SystemC provides a high-level-approach to modeling hardware behaviour and structure, rather than algorithms.

None of these approaches fully satisfy the requirements for pure RTL circuit design while using C-based languages, especially in terms of providing a consistent hardware, software, and concurrency model.

This work provides insights, in-depth examples, and evaluations of the Concurrent Programming and Processing (ConPro) HLS toolkit using, on the one hand, a traditional procedural programming language with some object-oriented features, and, on the other hand, a programming language extended with features enabling the composition of parallel systems based on the concurrent communicating sequential processes (CCSP) model. In contrast to other established HLS tools, ConPro requires no expert knowledge about hardware design, although it can be added if necessary. The ease of use of ConPro and its programming language is comparable with Arduino software frameworks. A simple "hello world" cicuirt requires less than 50 lines of code.

Another important requirement of a "hardware programming language" in circuit design (in contrast, to software design) is the ability to have fine-grained control over resource allocation, data types and ranges, and the synthesis process itself. ConPro provides fine-grained control that is necessary at the block level (parametrized blocks). Conpro generates structural VHDL, which can be synthesized to the gate level with common RTL synthesis tools such as Xilinx ISE or Synopsis Design Compiler.ConPro also generates design-specific synthesis make files for controlling the RTL synthesis process from the command line. Parallelism is explicitly modeled at the control-path and data-path levels.Data-path parallelism can be exploited by ConPro automatically, too.

This extended article gives an in-depth introduction to SoC-design methodologies using the rule-based High-Level Synthesis ConPro compiler framework and a process-oriented programming language. The Conpro programming language bases on the concurrent communication sequential processes (CCSP) model with an extensive set of interprocess-communication primitives. On one hand, this language is as a typical procedural sequential programming language like C or Modula, on the othe rhand, it supports parallel processing inherently and provides access to hardware components with only a few statements required. Conpro can be compared with simplicity of Arduino, allowing even beginners to design SoC designs, e.g., using FPGA prototyping technologies and widely used VHDL RTL synthesis software (e.g., Xilinx ISE Foundation adn WebPack).

Extended case studies of a smart communication protocol router and an advanced stack-based processor providing a programmatical virtualization layer are shown and evaluated. Both are used together as a smart node architecture deployed in high density sensor-actuator-networks, e.g., for material-integrated intelligent systems posing specific constraints.

The structure of this paper is as follows. First, an in-depth introduction to ConPro synthesis architecture and methods is given. The second main part is related to the ConPro programming language and gives an introduction with a lot of prominent examples. The third main part is dedicated to several extended case studies with evaluation that are presented. The appendix of this paper gives a compact tabular summary of the ConPro programming language.

ConPro was successfully deployed in lower and upper grade university courses to teach hardware design from high-level specifications **[EDU09]**. Even beginners with low experience in hardware design and hardware modelling were able to create advanced circuits.

# II ConPro SoC High-level Synthesis

Efficient hardware design requires more knowledge and configuration about components, protocols, and objects than classical languages like C can provide, for example, in terms of true bit-scaled registers, access, and implementation models at the architecture level (for example, single port versus dual port RAM blocks, static versus dynamic access synchronization). The generic software approach only covers the implementation of algorithms, but in hardware design, the synthesized circuit must be connected to and react with the outside world (other circuits, communication links, and many more), thus there must be a programming model to interface with hardware blocks, consistent with the imperative programming model. Furthermore, there must be a way to easily implement synchronization, which is always required in the presence of concurrency (at least on the control path level). A multiprocess model, established in the software developer community, provides a common approach for modelling parallelism, which is the preferred approach to implement and partition reactive systems on an algorithmic level.The ConPro programming language and synthesis **[BOS10A][BOS11A]** address the issues described above and introduce a design methodology for SoCs using the concurrent multiprocess model and the advanced behavioural programming language discussed in detail in the following Sections.

The synthesis flow is defined by a set of rules $\chi$ shown in Equation (*1.1*). Each set consists of subsets, which can be selected by parameter settings (for example, scheduling like loop unrolling, or different allocation rules) on the programming block level:

$$\text{Synthesis}: CP \xrightarrow{\chi^1} AST \xrightarrow{\chi^2} \begin{cases} \mu CODE \xrightarrow{\chi^3} \begin{cases} RTL \xrightarrow{\chi^4} VHDL \\ C/OCaML \end{cases} \\ C/OCaML \end{cases} \tag{1.1}$$

## II.1 Synthesis Flow

The processing architecture of the ConPro compiler and the synthesis flow are shown in **Fig. 1**. The synthesis process is a traditional software compiler flow with an intermediate representation (Synthesis Layer 1, preserving parallel processing and CSP-related features). There is a broad range of back-ends supporting software output, too (Synthesis Layer 3).

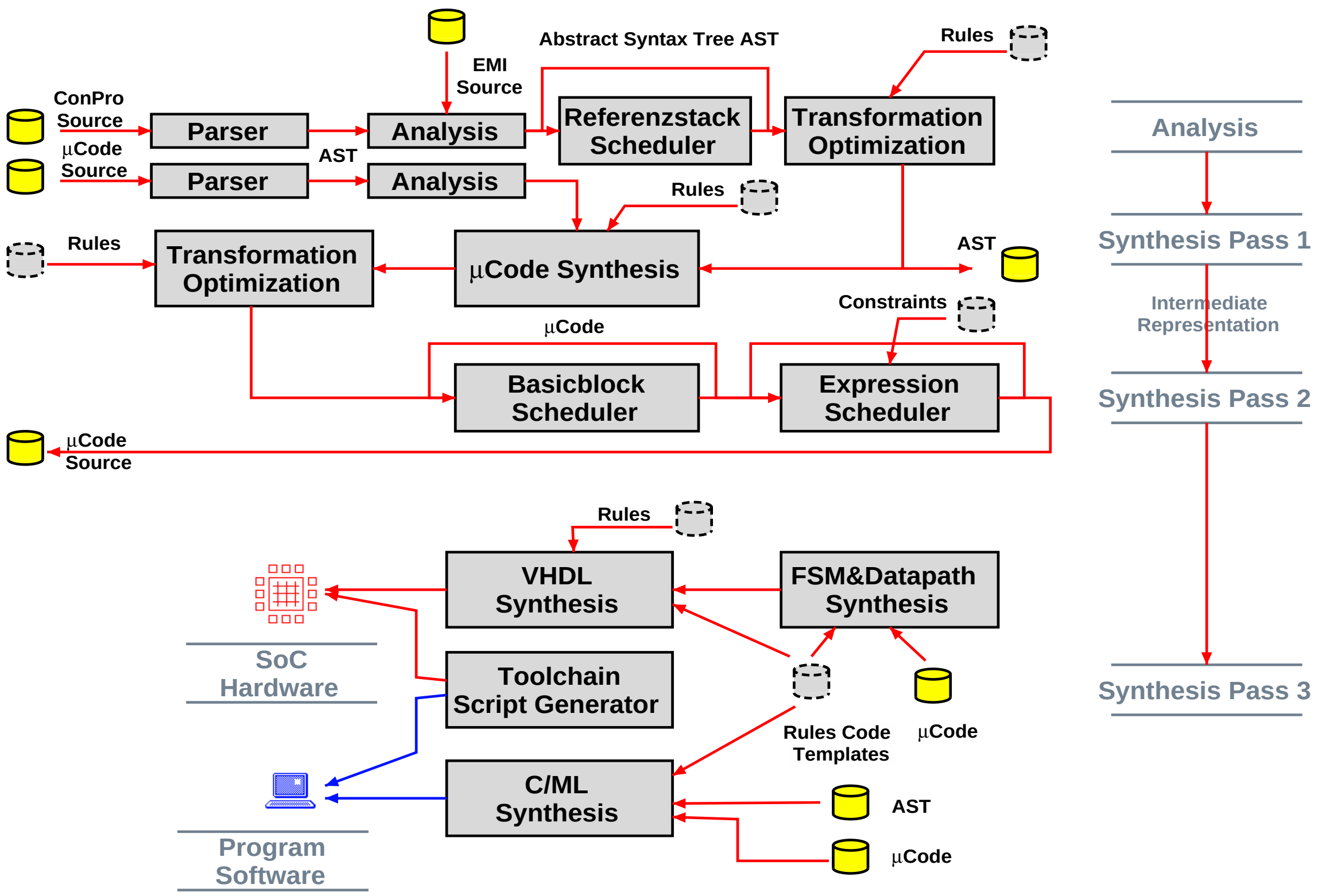

*Fig. 1* *Design flow using the high-level synthesis framework ConPro that maps the parallel CCSP programming model to SoC-RTL hardware and alternative software targets.*

The synthesis of RTL digital logic circuits from high-level *ConPro* program sources passes different phases:

1. First, the source code is parsed and analysed. For each process, an abstract syntax tree (AST) preserving complex statements is built. Global and local objects are stored in symbol tables (one globally at the module level, and one for each process level).First, optimizations are performed on the process instruction graph, for example, constant folding and dead object checking, and the elimination of those objects and superfluous statements. Several program transformations (based on rules and pattern matching) are performed, including temporary expression and register inference.A symbolic source code analysis method, called reference stack scheduler **[KU92]**, examines (local) data storage objects and their history in expressions. The reference stack scheduler analyses the evaluation of data storage expressions with an expression stack, one for each object.A sequence of storage assignments with expressions $\kappa=\{\Theta\leftarrow\varepsilon_1,\ \Theta\leftarrow\varepsilon_1,\ ..\}$ of a specific storage object $\Theta$ is transformed by the reference stack into a sequence of immutable and unique symbolic variables $\omega_i : \{\omega_1\leftarrow\varepsilon_1,\ \omega_2\leftarrow\varepsilon_1\ ,\ ..\}$. The aim is to reduce statements (using backward substitution and constant folding) and superfluous storage. The reference stack scheduler has an ALAP scheduling behaviour
2. After the analysis and optimization on the instruction graph level, the complex instructions (ranging from expressions to loops) of the AST are transformed into a linear list of µCode instructions. The µCode level is an intermediate representation of program code that is also used in software compilers, though no architecture-specific assumptions are made on this level other than control flow constraints.The µCode can be exported and imported, too. This feature enables a different entry level for other programming language front-ends, for example, functional languages **[SHA98]**.This intermediate representation allows for more fine-grained optimization, allocation, and scheduling. The transformation from the syntax graph to µCode infers auxiliary instructions and registers (suppose for-loops that require initialization, conditional branching, and loop-counter statements).Parallelism on the data path level is provided by a bind instruction that binds multiple instructions to one execution step or one FSM state (one time unit).
The transformation is based on a set of rules, $\chi_{\kappa\rightarrow\mu}$, consisting of default rules and user-selectable rules (constraint rules). This is the first phase of architecture synthesis, in which the source language paradigms are replaced with target machine paradigms, in this case a FSM with statements mapped on states and expressions mapped to the data path (RTL). In addition, the first phase of allocation is carried out here.Data path concurrency is explored either by user-specified bound blocks or by the basic block scheduler. This scheduler partitions the mCode instructions into basic blocks. These blocks have only one control path entry at the top and one exit at the tail. The instructions of one basic block (called a "major block") are further partitioned into minor blocks (containing at least one instruction or bound block). From these minor blocks, data dependency graphs (DDG) are built. Finally, the scheduler selects data-independent instructions from these DDGs with ASAP behaviour.

3. Following the first level of synthesis, the intermediate µcode is mapped onto an abstract state graph RTL using a set of rules $\chi_{\mu\rightarrow\Gamma\Delta}$, which includes both default and user-selectable rules. A final conversion step emits VHDL code. This design choice provides the possibility of adding other or new hardware languages, like Verilog, without changing the main synthesis path. The rule set determines resource allocation for temporary registers and functional blocks, providing different allocation strategies: shared versus non-shared objects, flat versus shared functional operators, and inference of temporary registers. Shared registers and functional blocks introduce signal selectors inside the data path.
The RTL units are partitioned into a state machine FSM (two hardware blocks, one transitional implementing the state register and one combinational implementing the state switch network), which provides the control path, and the data path (consisting of transitional and combinational hardware blocks, implementing functional operators, and access to global resources and local registers).
4. Using the default set of rules, each µCode instruction (except those in bounded blocks) is mapped to one state of the FSM requiring one time unit (≥ 1 clock cycle execution time, depending on object guards and process blocking). The rule set $\chi_{\kappa\rightarrow\mu}$, rather than $\chi_{\mu\rightarrow\Gamma\Delta}$, determines scheduling.

Although no traditional iterative scheduling and allocation strategies are applied in this compiler flow, the non-iterative constraint-selective and rule-based synthesis approach provides inherent scheduling and allocation with strong impact from different optimizers. To summarize, there are different levels of scheduling and allocation:

**Reference Stack Scheduler**
: Operates on the syntax graph level and tries to reduce statements, functional operators, and storage, and has an impact on scheduling and allocation.

**Basic Block Scheduler**
: Operates on the intermediate mCode level and tries to reduce operational time steps of statements, but has only an impact on scheduling.

**Expression Scheduler**

To satisfy timing and longest combinational path constraints, mainly clock-driven, complex, and nested flat expressions must be partitioned into sub-expressions using temporary registers and expanded scheduling. This scheduler has an impact on both scheduling and allocation.

**Optimizer**

Classical constant folding, dead code and object elimination, and loop/branch-invariant code transformations further reduce time steps and resources (operators and storage).

**Synthesis Rules**

The largest impact on scheduling and allocation comes from the set of synthesis rules $\chi = \chi_{\kappa\rightarrow\mu} \cup \chi_{\mu\rightarrow\Gamma\Delta}$

## II.2 Microcode Intermediate Representation

ConPro source code is parsed and analysed in the first compilation stage into symbol lists and syntax graphs of complex instructions or environments for each process. In the secondary compilation stage, the process instruction graph structure is flattened into a linear list of microcode program instructions (µCODE). The intermediate representation can be directly mapped on RTL using VHDL and Verilog models, partitioning the linear list of µCODE into a control and a data path. A more general form of µCODE is the generic and more natural RTL language µRTL, discussed in Sec. **III**. µCODE is the instructional assembler version of µRTL.

The main advantage of this approach is the simplified processing of a linear list of simple instructions from a small set rather than a complex graph structure with different complex programming environments and control statements, for example, for-loops. Furthermore, this microcode approach allows for the synthesis and implementation of microprocessor cores with a more traditional hardware/software co-design, which is not considered here but allows for future extensions of the synthesis flow.The set of µCODE instructions is summarized in **Tab. 1**.

The first two instructions relate to the data path, the jump instructions relate to the control path. The function application instruction is required for the access of Abstract Data Type Objects (ADTO) and their implementation. The ADTOs are commonly global resources that are shared by multiple processes. They require mutual exclusion synchronization and can provide higher level interprocess communication.

*Tab. 1 Language syntax of µCODE instructions: (Left) instructions, (Right) operand formats*

| Instruction | Description | Operand | Description |
|---|---|---|---|
| `MOVE(dst,src)` | Data transfer<br>dst ← src | `$immed.[i]` | Temporary symbolic variable used immediately in the next instruction(s) |
| `EXPR(dst, op1, operation, op2)` | Evaluate expression and assign result to `dst`. | `$temp.[i]` | Temporary storage register |
| `BIND(N)` | Bind the following *N* instruction to one bounded block. | `$alu.[i]` | Shared ALU |
| `JUMP(target)` | Branch to program position `target` | `DT{DT[N]}` | Data type |

| Instruction | Description | Operand | Description |
|---|---|---|---|
| `FALSEJUMP( cond:target)` | Branch to program position `target` if `cond` is false | `ET{`*DT*`[`*N*`]}` | Expression data type |
| `FUN(name⟦,arg⟧)` | Object operation or function application | `TC{`*DT*`[`*N*`]}` | Data type conversion |
| *label*`:` | Symbolic instruction (branch target) label | | |
| `SPECIAL(` *instruction*`)` | Internal usage | | |
| `NOP` | No operation place holder | | |

The special and label instructions are auxiliary instructions. The bind instruction provides a way to bind data path instructions that should be executed concurrently, which can be either set explicitly on the programming level or implicitly derived by a scheduler and optimizer exploiting parallelism at the data path level.

One obvious disadvantage of this approach is the loss of parallelism inside a process block, because each instruction or operation occupies at least one time step. To preserve parallelism at the process instruction level, the bind instruction was introduced. This instruction groups the *N* following data path instructions, and all instructions in the group are executed concurrently in one time step (a higher number of clock cycles may be required due to guarding and blocking constraints).

Operands of µCODE instructions can be registers, variables, signals, or temporary (symbolic) variables with additional information about data (DT) and expression types (ET) attached (sub-typing), required for type and data width conversion (TC) of operands to target object (LHS) types. More information is available about locality and LHS (write reference) or RHS (read reference) attributes, and if a shared ALU is used for an expression evaluation. The µCODE instructions provide initial scheduling and allocation information for the following RTL synthesis, including expression transformations. A short example of µCODE generation is shown in **Ex. 2**. The µCODE assembler module related to one process is partitioned into import, data, and code segments.

Ex. 2 µCODE synthesis examples

```
reg d : logic[8];
process foo:
begin
  reg a,b : logic[8];
  d <- 0;
  for i = 1 to 100 do
  begin
    a <- d, b <- d+1;
    d <- i + a + b;
  end
end;
----------------------- µCODE
import:
begin
  register d: L8
end

data:
```

```
begin
  register LOOP_i_0: I9
  register a: L8
  register b: L8
end

code:
begin
          i1_assign:
                     move (d,0)
      i1_assign_end:
                     nop
        i2_for_loop:
                     move (LOOP_i_0,1)
   i2_for_loop_cond:
                     bind (2)
                     expr ($immed.[1],100,>=,LOOP_i_0)
                     falsejump ($immed.[1],%END)
       i3_bind_to_4:
                     bind (3)
                     move (a,d)
                     expr (b,d,+,1)
                     nop
   i3_bind_to_4_end:
                     nop
          i5_assign:
                     bind (2)
                     expr ($immed.[2],LOOP_i_0:L0,+,a)
                     expr (d,$immed.[2],+,b)
                     nop
      i5_assign_end:
                     nop
   i2_for_loop_incr:
                     bind (3)
                     expr (LOOP_i_0,LOOP_i_0,+,1)
                     nop
                     jump (i2_for_loop_cond)
    i2_for_loop_end:
end
```

## II.3 Synthesis Rules

The simplest core set of synthesis rules applied to source code in the HLS is summarized below.

**Rule P: Process**

Each sequential *ConPro* process is mapped to one RTL block, with a FSM offering the control flow of the computation and a behaviour given by a state-transition-graph $\Gamma=(\Sigma,\psi)$, and a data path $\Delta$. The set of control states is $\Sigma=\{\sigma_1, \sigma_2, ..\}$, and the set of control state transitions is $E$. The entire RTL block is composed of combinational and transitional hardware blocks, $HW=(\Phi \cup \Pi)$.

**Rule SI: Storage**

There are only registers $\Re$ with independent data widths in the range $[1..W_{max}]$ bits.

```
reg name: DT[width];
```

**Rule SII: Storage Resource Sharing and Allocation**

There is no resource sharing: For each register $r \in \Re$ a transitional hardware block $\tau$ is allocated.

FUN ALLOCATE$^{STORAGE}$ : $\{r_1, r_2, .. \mid r_i \in \Re\} \rightarrow \{\tau_1, \tau_2, .. \mid \tau_i \in \Pi\}$

**Rule SIII: Statement Scheduling**

Each instruction $i$ from the set of elementary statements $\kappa$ is mapped to one state $\sigma$ of the control flow $\Gamma=(\Sigma,\psi)$ of the FSM. The set of elementary statements consists of $\kappa$ = {*Assignment*, *Expression Evaluation*, *Conditional Branch*, *Unconditional Branch*}.

FUN SCHEDULE : $\{i_1, i_2, .. \mid i_i \in \kappa\} \rightarrow \{\sigma_1, \sigma_2, .. \mid \sigma_i \in \Sigma\}$

**Rule SIV: Data path**

The data path $\Delta$ is decomposed and partitioned according to the control state set $\phi$.

**Rule SV: Expression Resource Sharing and Allocation**

There is no resource sharing. For each functional operator $op \in \aleph$ a combinational hardware block $\phi$ is allocated. The set of operators consists of $\aleph$={`+,-,*,/,and, or, not, shiftl, shiftr`}.

FUN ALLOCATE$^{EXPR}$ : $\{op_1, op_2, .. \mid op_i \in \aleph\} \rightarrow \{\phi_1, \phi_2, .. \mid \phi_i \in \Phi\}$

**Rule SVI: Complex Statements**

Complex statements $i \in \varepsilon$, for example, loops, are decomposed in a sequence of elementary statements using a synthesis decomposition rule $\chi$: $\varepsilon_i \rightarrow \{i_1, i_2, .. \mid i \in \kappa\}$. The set of complex statements consists of $\varepsilon$={*If-Then-Else*, *Case-Select*, *For-Loop*, *While-Loop*, ..}.

The transformation rules for some complex statements are shown in **Tab. 2** below.

*Tab. 2* *Synthesis transformation for complex ConPro statements. Notation is B: statement blocks, μCODE: x ← y; ➡ MOVE(y,x); eval(ε?), falsejump L; ➡ BIND(2); EXPR(t,ε); FALSEJUMP(t,L)*

| Statement | Transformation |
|---|---|
| for $cn_i$ = a to b do $B_i$ | ALLOCATE: $cn_i$ ➡ $r_i$ ➡ $\tau_i$<br>TRANSFORM: $cn_i$ ← a;<br>$LOOP_i$: eval($cn_i$ < b)?,<br>falsejump $EXIT_i$;<br>$B_i$;<br>$cn_i$ ← $cn_i$ + 1;<br>jump $LOOP_i$;<br>$EXIT_i$: ... |

| Statement | Transformation |
|---|---|
| if $e_i$ then $B_{i,1}$ else $B_{i,2}$ | TRANSFORM: EVAL($e_i$)?,<br>falsejump $LB_{i,2}$;<br>$B_{i,1}$;<br>jump $EXIT_i$;<br>$LB_{i,2}$: $B_{i,2}$;<br>$EXIT_i$: ... |
| match $e_i$ with<br>when $v_1$ : $B_{i,1}$<br>when $v_2$ : $B_{i,2}$<br>...<br>when others : $B_{i,n}$ | TRANSFORM: eval($e_i$=$v_1$)?,<br>falsejump $LB_{i,2}$;<br>$B_{i,1}$;<br>jump $EXIT_i$;<br>$LB_{i,2}$: eval($e_i$=$v_2$)?,<br>falsejump $LB_{i,3}$;<br>$B_{i,2}$;<br>jump $EXIT_i$;<br>...<br>$LB_{i,n}$: $B_{i,n}$;<br>$EXIT_i$: ... |

**Rule SVII: States and Transitions**

Each state $\sigma_i$ has always a successor state $\sigma_j$ with j≠i, except j=i if the current statement execution is blocked by a guard of a global object (satisfaction of the guard condition). A state transition is either conditional, given by the state transition function COND($\sigma_i$, $\sigma_j$ , *cond*) satisfying a precondition *cond*, or unconditional given by the function NEXT($\sigma_i$, $\sigma_j$) (execution of an ordered statement sequence):

FUN TRANSITION: $(\sigma_i,\sigma_j) \rightarrow \{\rho \mid \rho \in \{\text{NEXT}(\sigma_i, \sigma_j), \text{COND}(\sigma_i, \sigma_j, cond)\}\}$

There is a start and an end state ($\sigma_0$, $\sigma_\infty$) assigned to each sequential process. There is at least one state transitions to another state outgoing from each state, except for the end state $\sigma_\infty$, which has only an outgoing self transition. An example for a Conpro-µCODE transformation is shown in the following **Ex. 3**.

*Ex. 3* *µCODE synthesis from ConPro program snippet (nop instructions are removed during the µCODE optimization and state compaction)*

CONPRO

```
s <- 0;
for i = 1 to 10 do
begin
  t <- 1;
  s <- s + i;
  if s = 0 then
    t <- 2;
  end;
end;
```

MICROCODE

```
data:
        register LOOP_i_0: int[8]
code:
```

```
i1_assign:
  move (s,0)
i1_assign_end:
  nop
i2_for_loop:
  move (LOOP_i_0,1)
i2_for_loop_cond:
  bind (2)
  expr ($immed.[1],10, > =,LOOP_i_0)
  falsejump ($immed.[1],%END)
i3_assign:
  expr (s,s,+,LOOP_i_0:I8)
  nop
i3_assign_end:
  nop
i4_branch:
  bind (2)
  expr ($immed.[1],s,=,0)
  falsejump ($immed.[1],i2_for_loop_incr)
i4_branch_end:
  nop
i2_for_loop_incr:
  bind (3)
  expr (LOOP_i_0,LOOP_i_0,+,1)
  nop
  jump (i2_for_loop_cond)
i2_for_loop_end:
```

Advanced synthesis rules include basic block scheduling (on AST and *μCODE* level), reference stack based optimization (in AST level), and resource sharing (especially concerning temporary storage), discussed in the next Sections.

**Rule SOI: Extended set of storage objects**

The set of storage objects is extended with variables stored in RAM blocks and selected by their address

**Rule SOII: Resource sharing**

Inside processes, local storage like temporary registers can be reused for several computations. Furthermore, operations (arithmetic, relational, and boolean) can be shared with an ALU approach, using one or multiple ALU blocks.

**Rule SOIII: State compaction**

The set of control states and the state-transition graph of each process derived from statement sequences and complex statement decompositions can be reduced by merging control and data statements. This is mainly performed at the *μCODE* level.

**Rule SOIV: Optimization**

Several traditional optimization techniques like rescheduling (the Reference Stack approach), constant folding, and dead object/code optimization can improve SoC resource requirements and execution times significantly, including basic block parallelization, discussed in the next sections.

## II.4 Reference Stack (RS) Optimizer and Scheduler

The symbolic RS method offers storage and control flow optimization at the AST level using an automatic scheduling approach, which is divided into two passes:

1. Merging of expressions with ALAP scheduling behaviour on the abstract syntax tree level to resolve constant and register/variable folding beyond the initial instruction boundaries. The reference stack method merges expressions into meta expressions as large as possible, leading to optimised results in constant folding. On this level, all register and variable assignments are delayed as much as possible. Real data transfer with register and variable assignments is scheduled only on global block, branch, and loop boundaries.
2. Post scheduling of these (large) meta expressions with control step assignments depending on the chosen expression scheduling and allocation model on the intermediate microcode level. The meta expressions in a flat model are scheduled in a single time step. In the case of a two-ary or shared ALU model, the expression consisting of *N* two-ary operations is scheduled in N time steps with temporary register transfers.

Basically, the RS algorithm (explained in **Def. 1**) transforms a sequence of modifications of mutable storage objects into a sequence of immutable symbolic variables, as shown in **Ex. 4**. The flushing of storage assignments is postponed as much as possible. Control statements can trigger a flush depending on the occurrence of storage objects in conditional and loop body blocks. The algorithm in **Def. 1** is an advanced version adapted from **[KU92]**.

*Ex. 4* *Possible outcomes (right) of a source code block (left) using the reference stack analysis and optimization (case 2: y is not referenced after this block and has no side effects). Bottom: Transformation of mutable storage to immutable symbolic variables*

```
x ← a + b + 1;               ❶ y ← a + b + c;
y ← x + 1 + c;        ➡         x ← y;
x ← y - 2;                   ❷ x ← a + b + c;

DEF x0 = a + b + 1
DEF y0 = x0 + 1 + c
DEF x1 = y0 - 2              T(x)=[x1,x0]
```

***Def. 1*** ***RS Algorithm***

1. Each data object *x* is traced with its own reference stack *T*:

   $T(\mathrm{x})=[re_{\mathrm{i-1}};\ re_{\mathrm{i-2}};\ ...\ ;\ re_0]$

   (Left element is the top of the stack, *re* is a reference expression of type *stackelement*)

   Data objects x
   ```
   register, variable
   ```

   Reference Expressions
   ```
   TYPE stack = stackelement list
   TYPE stackelement =
     | RS_self(obj)
     | RS_expr(expr)
   ```

```
| RS_branch(stack list)
| RS_loop(stack)
| RS_ref(stackelement reference list)
```

2. An assignment

   $x_i \leftarrow \varepsilon(x_1, x_2, ...)$

   is delayed as late/last as possible (ALAP).
3. A new reference stack for object $x$ is created on the first occurrence of $x$ on LHS in an assignment:

   $x_i \leftarrow \varepsilon_0(x_1, x_2, ...)$ ➡ $T(x_i)$=[`RS_expr`($\varepsilon_0$)]

4. A new reference stack for object $x$ is created on the first occurrence of x on the RHS in an assignment or an expression:

   $T(x_i)$=[`RS_self`($x$)]

5. Each new assignment to a storage object $x$ updates the reference stack by pushing the new expression ε (RHS of assignment) on the stack $T(x)$.

   $x_i \leftarrow \varepsilon_n(x_1, x_2, ...)$ ➡ $T(x_i)$=[`RS_expr`($\varepsilon_n$), ..]

6. Each occurrence of a storage object in an expression is marked (in the AST) with a reference to the current top element of the reference stack for this element (with top=n-1 and n stack elements): $\uparrow T(x_1)_{top}$

   $\varepsilon_n(x_1, x_2, ...)$ ➡ $\varepsilon_n(x_1{:}\uparrow T(x_1)_{top}, x_2 : \uparrow T(x_2)_{top}, ...)$

   Inside branches or loops the `RS_ref`([*stack references*]) value is placed on the reference stack $T(x)$ for objects only referenced but not modified inside branches or loops, which is an important information for scheduling decisions (an RS flush, see next step).

7. Flush all delayed assignments at the end of the outer scheduling block and if there are no control statements like branches and loops before. After a flush has happened, the top of the reference stack is now `RS_self`. Resolve dependencies of objects and stack top expressions, sort assignments before flushing in correct dependency order.
   Relocate all references `RS_ref` pointing to other stack elements in expressions in the flushed assignments.

```
Example

x ← 12;        ✎ delay assignment, T(x)=[12]
y ← x + 1;     ✎ delay assignment, T(y)=[x0+1]
x ← v;         ✎ delay assignment, T(v)=[v], T(x)=[v0,12]
a ← y - 1;     ✎ delay assignment, T(a)=[y0-1]

➡ flush all delayed assignments with stack relocation and
  constant folding:

y ← y0=x0+1=12+1=13;
```

$a \leftarrow a_0=y_0-1=x_0+1-1=12;$
$x \leftarrow x_1=v_0=v;$

**8. Branches**

i. Evaluate all conditional blocks $B_{cond, i}$ of the branch, with total b different blocks, each for one branch condition, $j$=0...$b$-1. Each conditional block is handled with its own sub-stack $T_{Bcond,i}(x)$ with $n$ as the number of conditional blocks:

$T(x)$= [`RS_branch`([$T_{n-1}$;$T_{n-2}$,..,$T_0$],..]

All objects modified inside the conditional block $B_{cond, i}$ get a

`RS_branch`([$T_{Bcond,1}$= [], .., $T_{Bcond, i}$=[`RS_expr`(ε),..], ..])

stack element on the top of respective sub-stack. All further modifications are stored in this `RS_branch` stacks. Objects x appearing the first time in expressions get an `RS_branch`([.., $T_{Bcond, i}$=[↑$T(x)_{top}$], ..]) stack element. The initial top element $T_0$ of the branch is either a reference `RS_ref` to a previous expression or `RS_self`. The last element is required in loop environments due to side effects.

ii. Objects only referenced in each branch, that means they were not modified and $T_{Bcond,i}$=[`RS_self` | `RS_ref`] | ∉ `RS_expr` for all $i$=0...$b$-1, are transformed into references to previous stack elements:

$T(x)$=[`RS_branch`([[`RS_self`],[`RS_self`];...]),$T_{n-2}$ ,..]
➡
$T(x)$=[`RS_branch`([[`RS_ref`(↑$T_{n-2}$ )],[`RS_ref`(↑$T_{n-2}$ )];..]),$T_{n-2}$ ,..]

iii. After all conditional blocks were evaluated, objects with `RS_branch`([$T_{n-1}$; $T_{n-2}$;...;$T_0$]) and $T_i$≠ [`RS_self` | `RS_ref`] | ∈ `RS_expr` (modified inside the block) must be pushed before the branch instruction (expectation: all possible branches modify $x$) if there is an Expression `RS_expr` before the actual branch stack, and at the end of the conditional block they were modified.

iv. Modify the reference stacks:

a.) Storage object $x$ was not modified inside the branch, but referenced:

$T(x)$=[`RS_branch`([[`RS_self`],..]),`RS_expr`($ε_i$),..]
➡
$T(x)$=[`RS_ref`(`RS_expr`($ε_i$)),
    `RS_branch`([[`RS_ref`(`RS_expr`($ε_i$))], ..]),`RS_expr`($ε_i$), ..]

b.) $x$ was modified inside the branch:

$T(x)$=[`RS_branch`([[`RS_expr`($ε_j$),`RS_self`], ..]),`RS_expr`($ε_i$), ..]
➡
$T(x)$=[`RS_self`,
    `RS_branch`([[`RS_expr`($ε_j$),`RS_self`)], ..]),`RS_expr`($ε_i$), ..]

**9. Loops**

i. Evaluate all objects appearing in loop condition expressions before the loop body block is evaluated.
ii. Evaluate the body block *B* of the loop.
Objects appearing the first time in loop body expressions (and the RHS of assignments) get a `RS_loop`([$T_0$]) top stack element with $T_0$=`RS_self`. All objects modified the first time in the loop block get a `RS_loop`($T_{loop}$) expression on the top of stack. All further modifications are stored in this `RS_loop`($T_{loop}$) expression.

$T(x)$=[`RS_loop`($T_{loop}$=[..]), ..]

iii. Objects only referenced or appearing on the RHS in the loop body, that means $T_{loop}$=[`RS_self`] | ∉ `RS_expr`, are transformed in references to previous stack elements, and this reference is pushed to the stack, too:

$T$(X)=[`RS_loop`([`RS_self`]),$T_{n-2}$, ..]
➡
$T$(X)=[`RS_ref`(↑$T_{n-2}$), `RS_loop`([`RS_ref`(↑$T_{n-2}$)]),$T_{n-2}$, ..]

iv. After the loop block was evaluated, objects with `RS_loop`($T$) and $T\neq$ [`RS_self`] (modified inside the block) must be pushed before the loop instruction if there is an expression `RS_expr` before the actual loop, and inside at the end of the loop block, too.
v. Modify the reference stacks:
a.) Storage object *x* was not modified inside the loop body, but referenced:

T($x$)=[`RS_loop`([`RS_self`]),`RS_expr`($\varepsilon_i$), ..]
➡
T($x$)=[`RS_ref`(`RS_expr`($\varepsilon_i$)),
    `RS_loop`([`RS_ref`(`RS_expr`($\varepsilon_i$))]),`RS_expr`($\varepsilon_i$), ..]

b.) *x* was modified inside the loop body:

$T(x)$=[`RS_loop`([`RS_expr`($\varepsilon_j$),`RS_self`]),`RS_expr`($\varepsilon_i$), ..]
➡
$T(x)$=[`RS_self`,
    `RS_loop`([`RS_expr`($\varepsilon_j$),`RS_self`)]),`RS_expr`($\varepsilon_i$), ..]

## II.5 Basic Block Scheduling and Data Path Parallelization

The basic block scheduler explores the control flow and groups data path statements (assignments) in the largest possible statement blocks, the basic blocks. A basic block has only one control flow entry (the first statement) and only one exit (after the last statement). Basic blocks can be analysed regarding the exploitation of parallel execution of the statements of a basic block by using data dependency graphs (DDG). The DDG of a basic block is an acyclic directed graph with nodes representing the operations (assignments) and edges representing the data dependency of operations, determining an execution order.

A DDG of a basic block can indeed consist of a forest of independent graphs, and graphs with multiple root elements, as shown in **Fig. 2**.

Independent operations can be scheduled and executed in one cycle with a bound block. The goal is to schedule k operations in a bound block with $k$={1,..,$c$} and $c$ is either unlimited (maximal) or a limit of the maximal number of parallel operations. The DDG is divided into levels, which are then used for scheduling.

All operations belonging to one level are independent and can be scheduled in one bounded block, as shown in **Def. 2**. The level partition is performed by computing the longest path from a root node to each deeper node. Each node gets a marking of the longest path from a root node reaching this node, which is equal to the level.

It is assumed that the parallel execution of a set of instructions contained in a bounded block is atomic and each data transfer is only executed once.

```
(1)  t1  := a+1
(2)  t2  := a-1
(3)  y  := t1*t2
(4)  x  := t1
(5)  t3  := f(c)
(6)  z  := x-y
(7)  x  := x+1
(8)  t4  := t3-x
(9)  t5  := d
(10)  u  :=t5*t5
(11)  t6  := g(c)-1
(12)  u  := u-t6
```

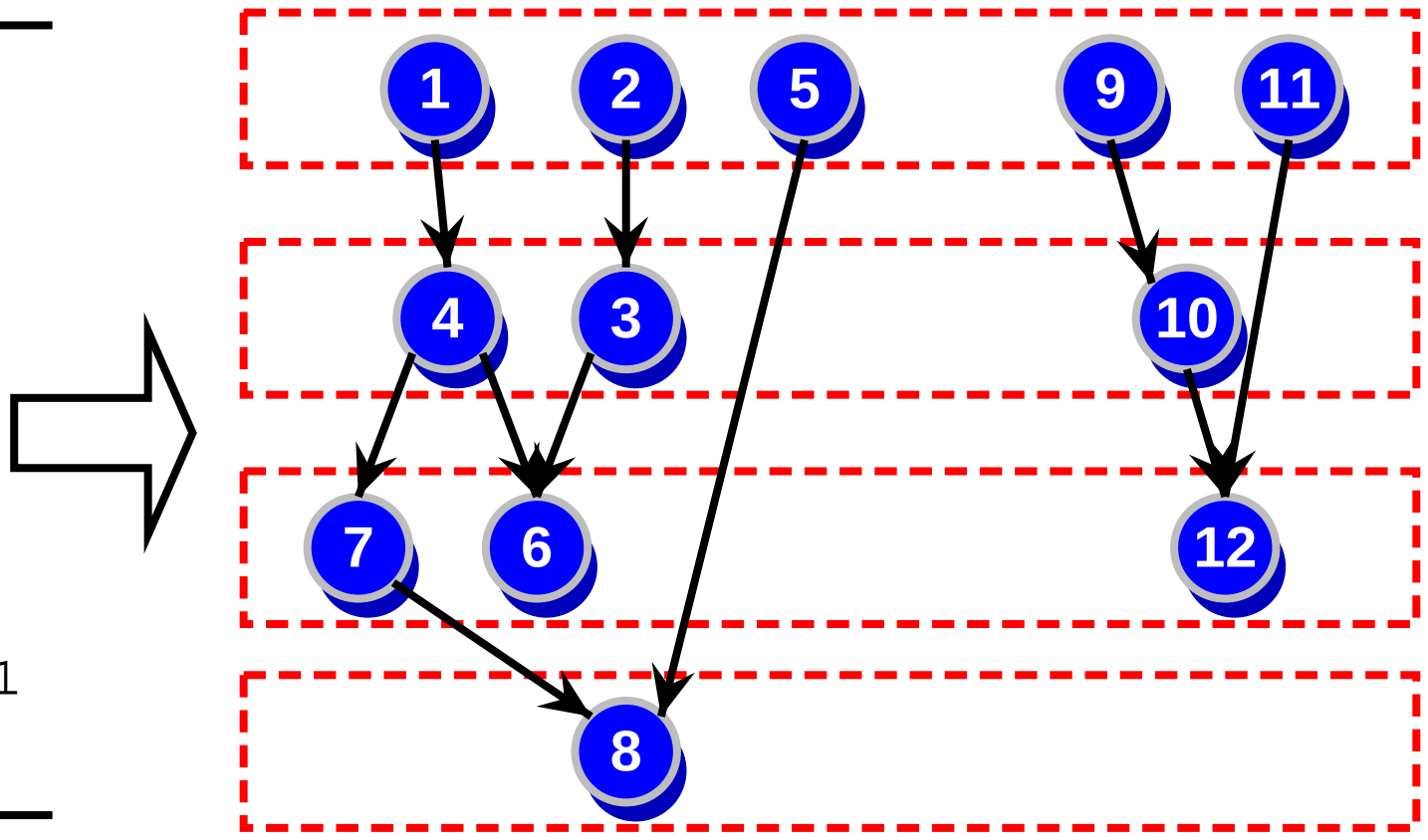

*Fig. 2* *The derivation of a DDG forest from a sequence of assignments and a possible parallel scheduling requiring only four bounded block statements (and cycles).*

That means that expressions on the RHS of all assignments in a bounded block are first evaluated (with the old values of all storage objects referenced in the expressions).

*Def. 2* *A.) Partition of an ordered list of instructions $K=\{i_1; ..\}$, with i from the set of statements κ, in a set of basic blocks $B=[b_1,b_2,..]$ consisting each of an ordered list of instructions (i;..). B.) The scheduling algorithm used to parallelize data flow independent sub-lists of instructions (i,..) mapped again to a flattened instruction list.*

```
FUN PARTITION: K=[i1;i2;..] → B=[b1;b2;..] IS
  b=[]; B=[];
  WHILE K ≠ [] DO
    k  ← HEAD(K), K ← TAIL(K);
    IF KIND(k) = DATA THEN b ← b ⊕ [k]
     ELSE B ← B ⊕ b, b ← [];


FUN SCHEDULE: B=[b1;b2;..] → K'=[(i1,i2,..);(..);..] IS
  K=[];
  ∀ b ∊ B DO
    DDG ← Build DDG forest from b;
    roots ← List of root nodes of DDG (nodes w/o predecessor);
    ∀ {op | op ∊ roots} DO
       Mark(op,1);
    maxLP ← 0;
    ∀ {op | op ∊ DDG\roots} DO
       LP ← 0;
```

```
        ∀ {op' | op' ∈ roots} DO
           LP ← Max(LP,|Path(op,op')|);
        Mark(op,LP);
        maxLP ← Max(maxLP,LP);
     ∀ {level | level ∈ {0..maxLP}} DO
        K ← K ⊕ [Bind({op ∈ DDG | Level(op) = level})];
```

## II.6 RTL Synthesis and VHDL Model

The synthesis of Register-Transfer-Level design is enabled by a transformation of the intermediate microcode instruction list representation to FSM and RT data path architecture blocks. The control and data paths of the microcode instruction list is represented using a (linear) state list. A state list consists of concatenated state expressions, which can be considered as control flow statements containing data flow expressions for the particular state:

A state expression of a state transition list consists of:

**state_name**

State name string in the format: `S_i`*<instrid>*`_`*<instrname>*`_`*<subop>*

**state_next**

Different control statements (Control path of the RTL):

**Next(*label*), Next_instr**

Control flow is directed to labelled state <*label*>. The next instruction kind is an unresolved pointer to the next following state expression in the list.

**Branch (*data list*, $next_0$, $next_1$)**

Conditional branch. Data contains the boolean expression required for evaluating the conditional branch. Up to two possible state transitions are possible.

**Select (*data list*, *case_state list*)**

Multi-case non-hierarchical branching. Data contains the expression(s) to which each case value(s) is (are) compared. [*case_state=(data list, next_state)*]

**state_data**

Specification of the data path of the RTL block, e.g., different data transfer statements, mainly handling *VHDL* signal assignments:

**Data_in( vhdl_*expr*)**

Input arguments for an expression, e.g., the RHS of a data transfer statement, for example operands for the ALU.

**Data_out (*vhdl_expr*)**

Output from an expression or operational unit is directed to the LHS of a data transfer statement, for example the ALU output result or the output from a global object, acting only as a resource multiplexer (part of the combinatorial data path).

**Data_trans(*vhdl_expr*)**

LHS of an expression, transitional local register data transfer (part of the transitional data path).

**Data_signal(*vhdl_expr*)**

Signal assignments (state dependent), static driven.

**Data_cond(*vhdl_expr*)**

Conditional expression (expression transformed *VHDL* If/Case).

**Data_top(*vhdl_expr*), Data_top_def (*vhdl_expr*)**

Entity top-level VHDL expressions and architecture definitions.

**Data_def(*vhdl_expr*,*vhdl_expr*), Data_def_trans (*vhdl_expr*,*vhdl_expr*)**

Default signal assignments in combinatorial and transitional data path.

**VHDL Output**

Finally, *VHDL* entities are created from the state transition list (there is one list for each sequential process). Additional *VHDL* entities are created for modules. Global objects are implemented in the module's top-level entity, shown in **Fig. 3**.

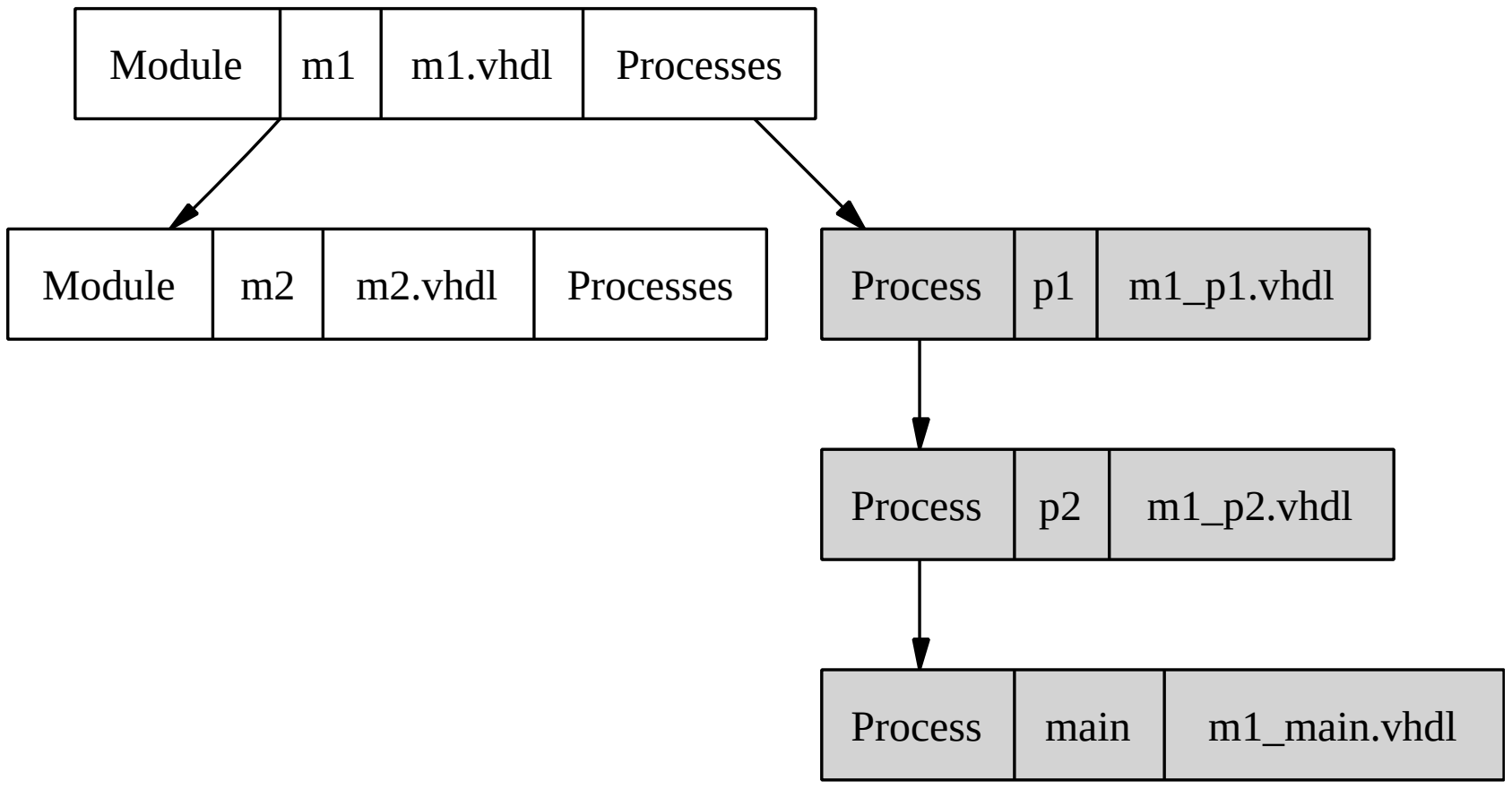

*Fig. 3* *ConPro module hierarchy and created VHDL components.*

The following example shows the outline of the synthesized *VHDL* from a small *ConPro* design consisting of three processes and two shared objects, a register array and a semaphore used to synchronize the three processes.

*Ex. 5* *A small ConPro design and excerpt of the synthesized VHDL*

ex.cp

```
open Core; open System; open Process; open Semaphore;

array data_in,data_out: reg[100] of int[16];
reg mon:int[16];           export mon;

object sem: semaphore with init=0;

function f(x:int[16]) return(y:int[16]): begin
  y <- x*x;
end;

process consumer1: begin
  sem.down();
  for i = 0 to 50 do begin data_out.[i] <- f(data_in.[i]); end;
  sem.up();
end;
```

```
process consumer2: begin
  sem.down();
  for i = 51 to 99 do begin data_out.[i] <- f(data_in.[i]); end;
  sem.up();
end;

process main: begin
  sem.init(0);
  for i = 0 to 99 do begin data_in.[i] <- i; end;
  consumer1.start(); consumer2.start(); sem.up();
  for i = 1 to 2 do begin sem.down(); end;
  for i = 0 to 99 do begin mon <- data_out.[i]; end;
end;
```

```
***********************************************************************
conpro ex.cp ➠
***********************************************************************
```

ex.vhdl

```
entity MOD_ex is
port(-- Connections to the outside world
  signal mon_RD: out std_logic_vector(15 downto 0);
  signal CLK: in std_logic;
  signal RESET: in std_logic
);
architecture main of MOD_ex is
  Process instances
  component ex_FUN_f
  port(..)
  component ex_consumer1
  port(..)
  component ex_consumer2
  port(..)
  component ex_main
  port(..)
  type ARRAY_data_out_TYPE is array(0 to 99) of
       signed(15 downto 0);
  signal ARRAY_data_out: ARRAY_data_out_TYPE;
  ..
  signal PRO_consumer1_ENABLE: std_logic;
  signal PRO_consumer1_END: std_logic;
  signal PRO_consumer1_main_START: std_logic;
  signal PRO_consumer1_main_GD: std_logic;
  ..
begin
```

ex_consumer1.vhdl

```
entity ex_consumer1 is
port(
  Connections to external objects, components and the outside
  world
  signal ARRAY_data_out_WR: out signed(15 downto 0);
  signal ARRAY_data_out_WE: out std_logic;
  signal ARRAY_data_out_GD: in std_logic;
  signal ARRAY_data_out_SEL: out std_logic_vector(7 downto 0);
  signal MUTEX_LOCK_FUN_f_LOCK: out std_logic;
  signal MUTEX_LOCK_FUN_f_UNLOCK: out std_logic;
```

```
  signal MUTEX_LOCK_FUN_f_GD: in std_logic;
  signal SEMA_sem_DOWN: out std_logic;
  signal SEMA_sem_UP: out std_logic;
  signal SEMA_sem_GD: in std_logic;
  signal REG_RET_FUN_f_y_RD: in signed(15 downto 0);
  signal PRO_FUN_f_CALL: out std_logic;
  signal PRO_FUN_f_GD: in std_logic;
  signal ARRAY_data_in_RD: in signed(15 downto 0);
  signal ARRAY_data_in_SEL: out std_logic_vector(7 downto 0);
  signal REG_ARG_FUN_f_x_WR: out signed(15 downto 0);
  signal REG_ARG_FUN_f_x_WE: out std_logic;
  signal PRO_consumer1_ENABLE: in std_logic;
  signal PRO_consumer1_END: out std_logic;
  signal conpro_system_clk: in std_logic;
  signal conpro_system_reset: in std_logic
);
end ex_consumer1;
architecture main of ex_consumer1 is
  Local and temporary data objects
  signal LOOP_i_0: signed(7 downto 0);
  State Processing
  type pro_states is (
    S_consumer1_start, -- PROCESS0[:0]
    S_i1_fun, -- FUN22215[ex.cp:19]
    S_i2_for_loop, -- COUNT_LOOP2877[ex.cp:20]
    S_i2_for_loop_cond, -- COUNT_LOOP2877[ex.cp:20]
    S_i3_fun, -- FUN78627[ex.cp:22]
    S_i4_assign, -- ASSIGN23890[ex.cp:22]
    S_i5_fun, -- FUN19442[ex.cp:22]
    S_i6_assign, -- ASSIGN58851[ex.cp:22]
    S_i7_fun, -- FUN83891[ex.cp:22]
    S_i2_for_loop_incr, -- COUNT_LOOP2877[ex.cp:20]
    S_i8_fun, -- FUN63052[ex.cp:24]
    S_consumer1_end -- PROCESS0[:0]
    );
  signal pro_state: pro_states := S_consumer1_start;
  signal pro_state_next: pro_states := S_consumer1_start;
..
  state_transition: process(..) begin
    if conpro_system_clk'event and conpro_system_clk='1' then
      if conpro_system_reset='1' or PRO_consumer1_ENABLE='0'
      then
        pro_state <= S_consumer1_start;
      else
        pro_state <= pro_state_next;
      end if;
    end if;
  end process state_transition;

  Process implementation
  Instruction Controlpath Block - The Leitwerk
  control_path: process(..) begin ..
    case pro_state is
      when S_consumer1_start => -- PROCESS0[:0]
        pro_state_next <= S_i1_fun;
      when S_i1_fun => -- FUN22215[ex.cp:19]
        if not((SEMA_sem_GD) = ('0')) then
          pro_state_next <= S_i1_fun;
        else
```

```
          pro_state_next <= S_i2_for_loop;
        end if;
      when S_i2_for_loop => -- COUNT_LOOP2877[ex.cp:20]
        pro_state_next <= S_i2_for_loop_cond;
      when S_i2_for_loop_cond => -- COUNT_LOOP2877[ex.cp:20]
        if CONST_I8_50 >= LOOP_i_0 then
          pro_state_next <= S_i3_fun;
        else
          pro_state_next <= S_i8_fun;
        end if;
      when S_i3_fun => -- FUN78627[ex.cp:22]
        if not((MUTEX_LOCK_FUN_f_GD) = ('0')) then
          pro_state_next <= S_i3_fun;
        else
          pro_state_next <= S_i4_assign;
        end if;
      when S_i4_assign => -- ASSIGN23890[ex.cp:22]
        pro_state_next <= S_i5_fun;
      when S_i5_fun => -- FUN19442[ex.cp:22]
        if PRO_FUN_f_GD = '1' then
          pro_state_next <= S_i5_fun;
        else
          pro_state_next <= S_i6_assign;
        end if;
      when S_i6_assign => -- ASSIGN58851[ex.cp:22]
        if ARRAY_data_out_GD = '1' then
          pro_state_next <= S_i6_assign;
        else
          pro_state_next <= S_i7_fun;
        end if;
      ..
      when S_consumer1_end => -- PROCESS0[:0]
        pro_state_next <= S_consumer1_end;
        PRO_consumer1_END <= '1';
end process control_path;

Instruction Datapath Combinational Block
data_path: process(..) begin
  -- Default values
  SEMA_sem_DOWN <= '0';
  MUTEX_LOCK_FUN_f_LOCK <= '0';
  REG_ARG_FUN_f_x_WR <= to_signed(0,16);
  REG_ARG_FUN_f_x_WE <= '0';
  ARRAY_data_in_SEL <= "00000000";
  PRO_FUN_f_CALL <= '0';
  ARRAY_data_out_WE <= '0';
  ARRAY_data_out_SEL <= "00000000";
  ARRAY_data_out_WR <= to_signed(0,16);
  MUTEX_LOCK_FUN_f_UNLOCK <= '0';
  SEMA_sem_UP <= '0';
  case pro_state is
    when S_consumer1_start => -- PROCESS0[:0]
      null;
    when S_i1_fun => -- FUN22215[ex.cp:19]
      SEMA_sem_DOWN <= SEMA_sem_GD;
    when S_i2_for_loop => -- COUNT_LOOP2877[ex.cp:20]
      null;
    when S_i2_for_loop_cond => -- COUNT_LOOP2877[ex.cp:20]
      null;
```

```
        when S_i3_fun => -- FUN78627[ex.cp:22]
          MUTEX_LOCK_FUN_f_LOCK <= MUTEX_LOCK_FUN_f_GD;
        when S_i4_assign => -- ASSIGN23890[ex.cp:22]
          REG_ARG_FUN_f_x_WR <= ARRAY_data_in_RD;
          REG_ARG_FUN_f_x_WE <= '1';
          ARRAY_data_in_SEL <= (I_to_L(LOOP_i_0));
        when S_i5_fun => -- FUN19442[ex.cp:22]
          PRO_FUN_f_CALL <= '1';
        when S_i6_assign => -- ASSIGN58851[ex.cp:22]
          ARRAY_data_out_WR <= REG_RET_FUN_f_y_RD;
          ARRAY_data_out_WE <= '1';
          ARRAY_data_out_SEL <= (I_to_L(LOOP_i_0));
        ..
    end process data_path;

    Instruction Datapath Transitional Block
    data_trans: process(..) begin
      if conpro_system_clk'event and conpro_system_clk='1' then
        if conpro_system_reset = '1' then
          LOOP_i_0 <= to_signed(0,8);
        else
          case pro_state is
            when S_consumer1_start => -- PROCESS0[:0]
              null;
            when S_i1_fun => -- FUN22215[ex.cp:19]
              null;
            when S_i2_for_loop => -- COUNT_LOOP2877[ex.cp:20]
              LOOP_i_0 <= CONST_I8_0;
            ..
            when S_i2_for_loop_incr =>
              -- COUNT_LOOP2877[ex.cp:20]
              LOOP_i_0 <= LOOP_i_0 + CONST_I8_1;
```

ex_consumer2.vhdl

```
          entity ex_consumer2 is ..
```

ex_FUN_f.vhdl

```
          entity ex_FUN_f is
          port(
            -- Connections to external objects, components and the outside world
            signal REG_RET_FUN_f_y_WR: out signed(15 downto 0);
            signal REG_RET_FUN_f_y_WE: out std_logic;
            signal REG_ARG_FUN_f_x_RD: in signed(15 downto 0);
            signal PRO_FUN_f_ENABLE: in std_logic;
            signal PRO_FUN_f_END: out std_logic;
            signal conpro_system_clk: in std_logic;
            signal conpro_system_reset: in std_logic
          );
          ..
```

ex_main.vhdl

```
          entity ex_main is
          ..
```

# III The μRTL Programming Language

In the previous section, the intermediate representation using micro-code instruction was introduced. These operations and declarations represent directly the RTL, which can be finally synthesized to gate-level using VHDL or Verilog models. The parallel programming language *Conpro* maps sequential processes onto RTL process blocks consisting of a:

1. Finite-State Machine (FSM) implementing the control path; and
2. Register-transfer implementing data paths.

Multi-process systems are implemented with interconnected multi-FSM data path structures. The behaviour of an RT process block is specified with the *μRTL* language introduced separately in this section (based on **[BAR73]**). This language can be seen as an independent core language to model and understand the behaviour of the extended concurrent CSP programming model. The syntax of the basic μRTL operations are summarized in **Tab. 3**. In contrast to the previously introduced μCODE assembler language, which is limited to one process machine, μRTL can be used to describe multi-process systems, too.

*Tab. 3 Minimal symbolic parallel programming language that can be immediately synthesized to hardware RT architectures (ε: computational expression).*

| Operation | Description |
|---|---|
| $r \leftarrow expr$ | Assignment of the value to a register. |
| $i_1$ ; $i_2$ ; $i_3$ ; ... | Sequential execution of statements. |
| $i_1$ , $i_2$ , $i_3$ , ... ; | Parallel execution of statements (limited to non-blocking statements, i.e., assignments) |
| branch *s* | An unconditional branch to a different instruction related to the state *s* before or after this statement. |
| branch *cond*:*s* | Conditional branch if the condition is satisfied, otherwise the next instruction is executed. |
| *s*: | An instruction label assigning a state to an instruction. |
| $o \rightarrow op(arg_1, arg_2, \ldots)$ | Procedural application of an operation to a global shared object with optional arguments. |
| $r \leftarrow \varepsilon(o \rightarrow op(arg_1, arg_2, \ldots))$ | Functional application of an operation to a global shared object with optional arguments returning a value. |
| process $p \rightarrow$ { $i_1$ ; $i_2$ ; $i_3$ ; ... } | Definition of a named process composition. A process *p* is treated as an object, too, supporting start and stop operations. |
| channel o<br>$o \rightarrow$receive \| send<br>join($p_1$, $p_2$, ..) | Some synchronisation primitives. The join operation can be used to temporally synchronize the control paths of different process (barrier). |

An example deriving the µ*RTL* specification from the *OCCAM* and Conpro programming language is shown in **Ex. 6**.

*Ex. 6 A µRTL specification derived from an OCCAM and ConPro program*

```
OCCAM

  [100] INT Data:
  INT Mean:
  SEQ
    Mean := 0
    SEQ Index = 0 FOR 100
      Mean := Mean + Data[Index]
    Mean := Mean / 100

ConPro

  array Data : reg of int[16];
  process main:
  begin
    reg mean: int[20];
    mean <- 0;
    for i = 1 to 100 do
    begin
      mean <- mean + Data.[i];
    end;
    mean <- mean / 100;
  end;

µRTL

process main → {
  s1: Index ← 0, Mean ← 0;
  s2: branch (Index=100):s6;
  s3: Mean ← Mean + Data[Index];
  s4: Index ← Index + 1;
  s5: branch s2;
  s6: Mean ← Mean / 100;
}
```

The relationship of the *µRTL* instruction to process algebra and the process flow behaviour of the µ*RTL* instructions is given below (♦: Synchronisation and scheduling point).

**Process Flow Behaviour**

| | |
|---|---|
| $r \leftarrow \varepsilon$ | ➡ $\{r \leftarrow \varepsilon\}$ ➡ ♦ |
| $i_1$ ; $i_2$ ; .. | ➡ $\{i_1\}$ ➡ ♦ ➡ $\{i_2\}$ ➡ .. |
| $i_1$ , $i_2$ , .. | ➡ $\{i_1\} \parallel \{i_2\}$.. ➡ ♦ |
| $o \rightarrow op()$ | ➡ $\{o \rightarrow op()\}$ ➡ $\Diamond \neg guard(op)$ ➡ |
| branch *cond*:*s* | ➡ ($\neg cond$ ➡ $\{i_{n+1}\}$ \| $cond$ ➡ $\{i_s\}$ ) ➡ ♦ |
| join (p1,p2, ..) | ➡ ($p_1 \parallel p_2 \parallel .. \parallel p_n$) ➡♦➡($p_1{}^* \parallel p_2{}^* \parallel .. \parallel p_n{}^*$) |

# IV The ConPro Programming Language

Concurrency is explicitly modelled at the control path level, with processes sequentially executing a set of instructions, initially independent of any other process.Parallel composition is available on the programming level but is static at run-time to enable the synthesis of SoC designs with static resources. That means the number of processes is fixed at design and compilation time.

Inter-Process Communication (IPC) provides process synchronization with different objects (Mutex, Semaphore, event, timer) and data exchange between processes using queues or channels, based on the CSP model. More fine-grained concurrency is provided on the data path level using bounded blocks executing several instructions (only on the data path, e.g., data assignments) in one time unit.

Block-level parallelism can be enabled explicitly or implicitly by a basic block scheduler.Two methods exist for accessing hardware blocks modeled on the hardware level (VHDL) from the programming level:

1. Using an object-orientated programming approach with methods
2. Using a signal component structure interface and signals

The first approach treats all hardware blocks, including IPC, as abstract data type objects (ADTO) with a defined set of methods accessible on the process level (at run-time) and on the top-level (only applicable with configuration methods, for example, setting the time interval of a timer). The bridge between the hardware and ConPro programming model is provided by the External Module Interface (EMI).

A signal component structure can be used to instantiate and access external hardware blocks and to create the top-level hardware port interface of the ConPro design. Signals are interconnection elements without a storage model. They provide only an interface to external hardware blocks. Signals are used in component structures, too.

The ConPro programming model is a common source for hardware and, alternatively, software synthesis, reusing the same program source for different implementations, as shown in Figure *4* on the left side. The EMI programming paradigm is a central part of this multi-target implementation synthesis by encapsulating and abstracting hardware blocks, for example, IPC and communication objects. Most EMI objects can be specified by a hardware (VHDL) behaviour and an operational equivalent and complementary software model without compromising the programming level, as shown in Figure *4* on the right side.

In **Fig. 4** the ConPro programming language block structure and the composition of top-level module designs using these blocks are summarized.

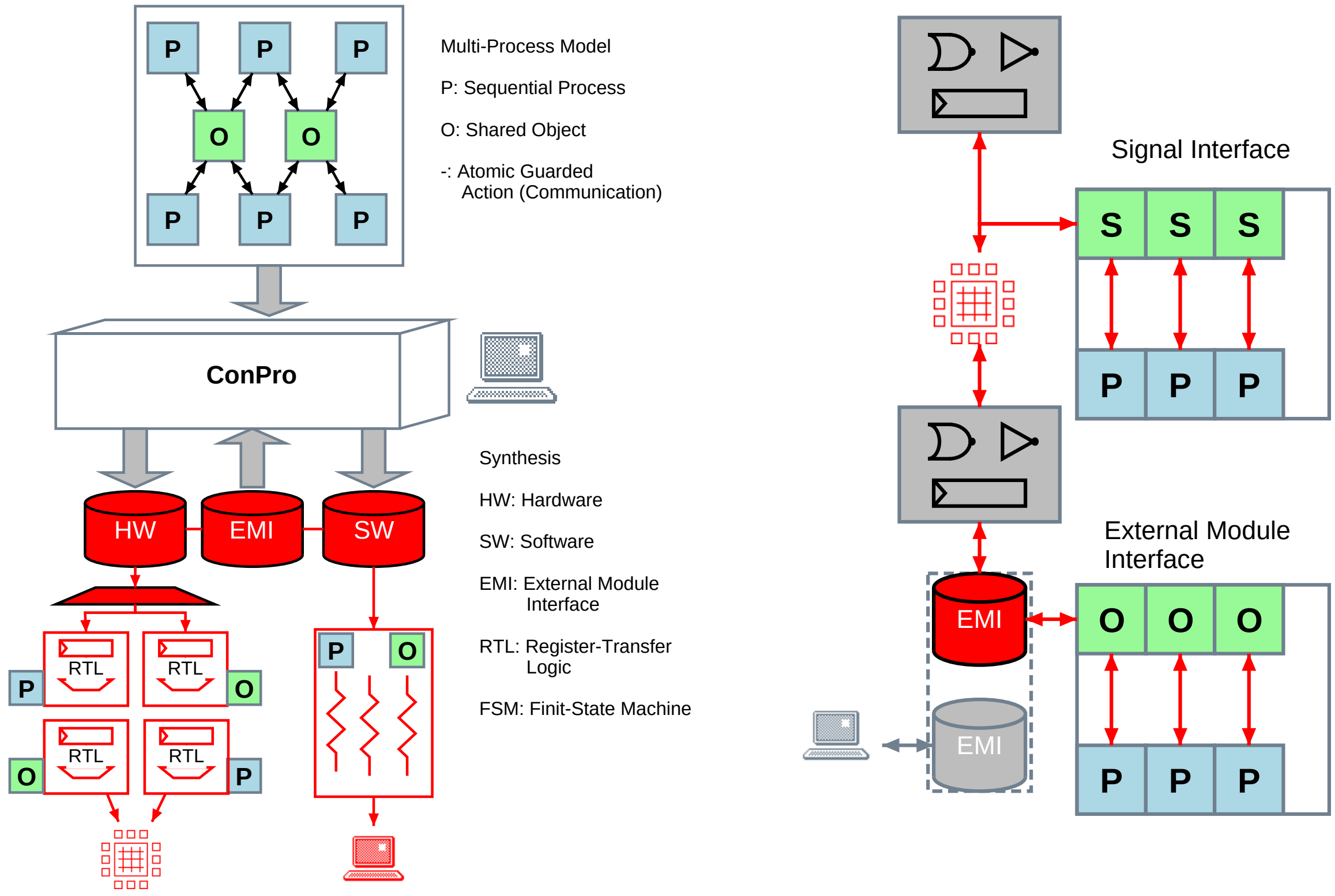

*Fig. 4* *Left: The ConPro programming model as a common source for hardware and alternatively software synthesis. Right: Hardware Interfaces from programming level, (a) Signal Interface, (b) EMI*

## IV.1 Hello World

For a first impression, let us start with the hello world demonstration. The entire ConPro code is shown in **Ex. 3**. The first five lines open core, system, and EMI modules. Core and Process modules are always required. The Timer module provides timer synchronisation objects, and the Uart module provides a serial link communication device accessed by a set of operations. For using frequency and time unit values in expressions, the clock frequency must be defined (line 9) via a system object created in line 8. The IO ports are defined and exported in lines 15-22. The top-level module is mapped to a component object (line 21), basically defining the IO ports by a separate structrue type definition in lines 15-19.

The main process (must always be defined) implements a time-synchronised rotation LED flashlight (lines 69-81). As in microcontroller software, the main process implements an endless loop serving the LED flashlight. The process control path is blocked by calling the timer await operation (line 77). The timer interval is set in line 37 (at compile time). The LED signals (output of the circuit) are directly connected to a register (mapping done in line 28), modified with bit operations in line 79.

The second part of the hello world circuit is a full serial receiver and sender via a common UART link with a receive and transmit line. There is a dedicated receiver process, defined from lines 40 to 66. The process uses a conditional loop to wait for one character received by calling the UART object read operation (line 51). This operation blcoks the control path of this process until either a character was received or an error occurred. After a character was received and stroed in the local process register *d*, the string `hello` will be sent.

The entire design compiles to 1311 lines of VHDL code (split into four source files, one for each process, the top-level component gluing all together, and a ConPro library file).

*Ex. 7 Hello world circuit with rotating LED patterns and UART text transmission.*

```
open Core;
open Process;
open Timer;
open System;
open Uart;

-- Hardware system parameters
object sys : system;
  sys.clock  (18500 kilohz);
  sys.target ("xc3s1000-ft256-5");
  sys.target ("xc18v04");
  sys.reset_level (0);

-- IO port type
type dev_type : {
  port leds: output logic[4];
  port rx: input logic;
  port tx: output logic;
};
-- IO port of FPGA
component DEV: dev_type;
export DEV;

-- LEDs
reg stat_leds,stat_ev: logic[4];
reg diag: logic[3];
-- Map register on output port
DEV.leds << stat_leds;

-- UART
object myuart : uart;
  myuart.interface(DEV.rx,DEV.tx);
  myuart.baud(115200);

-- LED toggle timer
object ledtimer: timer;
  ledtimer.time(500 millisec);

-- UART handler
process receiver:
begin
  reg d: logic[8];
  reg err : bool;
  myuart.init ();
  myuart.start ();

  err <- false;
  while err = false do
  begin
```

```
    -- wait for character
    myuart.read(d,err);
    -- send "hello"
    for i=1 to 5 do
    begin
      match i with
      begin
        when 1: d <- 'H';
        when 2: d <- 'e';
        when 3: d <- 'l';
        when 4: d <- 'l';
        when 5: d <- 'o';
      end;
      myuart.write(d,err);
    end;
  end;
end;

-- Main process
process main:
begin
  ledtimer.init  ();
  ledtimer.start ();
  receiver.start ();
  stat_leds <- 0b0001;
  always do
  begin
    ledtimer.await ();
    -- LED bit rotation
    stat_leds <- stat_leds[1]@stat_leds[2]@stat_leds[3]@stat_leds[0];
  end;
end;
```

## IV.2 Process Composition

Processes are the main execution units of a ConPro design module. A process executes a set of instructions in sequential order. Processes communicate with each other and the environment by using shared objects (inter-process communication, IPC). There is sequential and parallel process composition on the control and data path levels using bounded blocks, shown in **Fig. 5**.

Parallel process composition at the control path level is flat and provides only coarse-grained concurrency.

Parallel process composition on the data path level inside a process is fine-grained but limited to non-blocking statements not affecting the control path of a process.The ConPro programming language supports named processes defined at the top-level only, in contrast, for example, to the OCCAM programming model, which allows embedded and nested parallel and sequential process constructors anywhere.

A process definition consists of a header defining the unique process name and the process body defining local object definitions (types, data, and some abstract objects) and a sequence of statements.

Processes are abstract objects, too. Thus, there is a set of methods {`start`, `stop`, `call`} that can be applied to process objects (identifiers). They are controlling the state of a process.

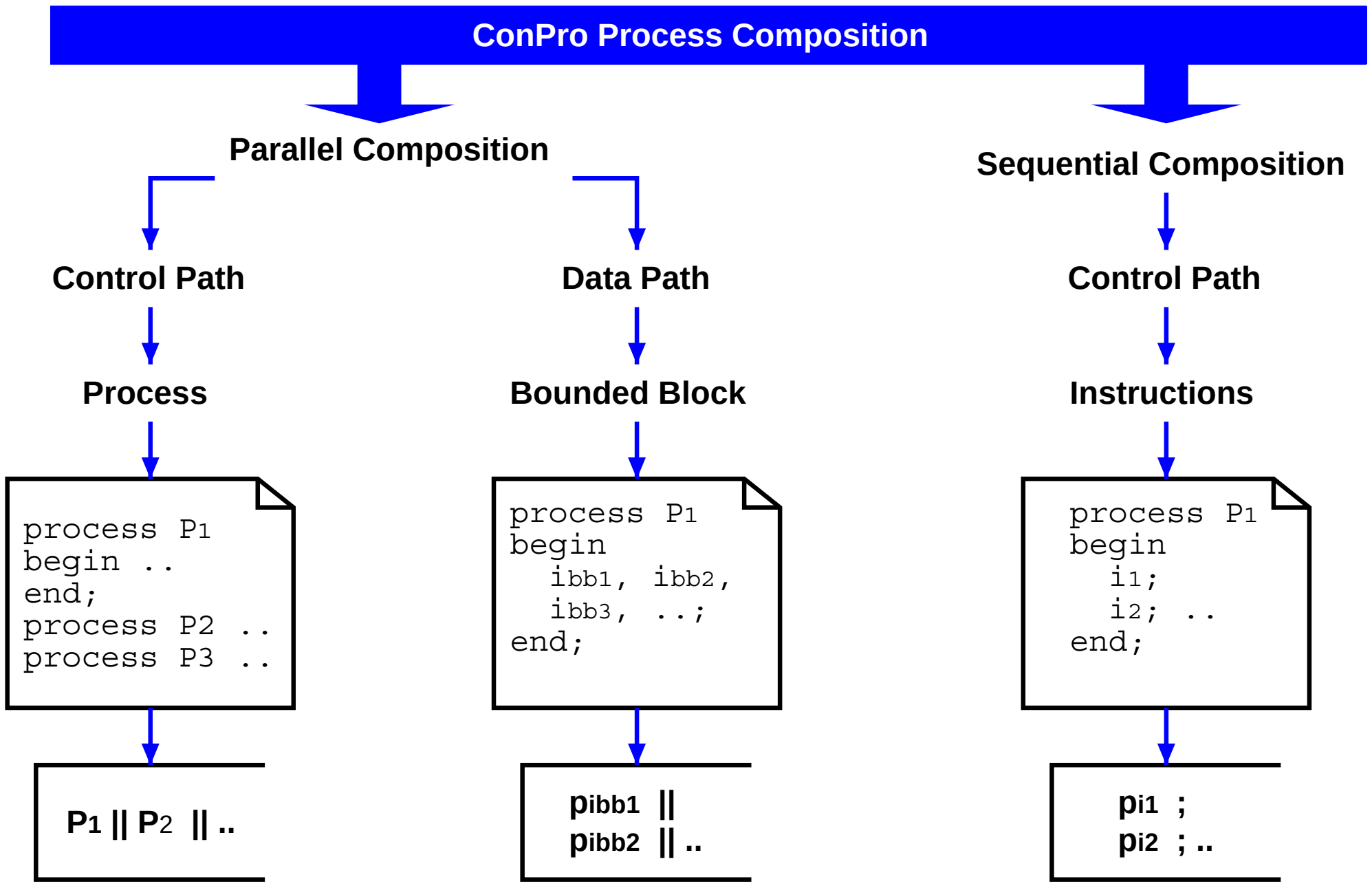

*Fig. 5* *Different levels of concurrency and process composition in ConPro*

A process must be explicitly started by another process by using the `start` method, which starts the process concurrently (asynchronous call). A process can be started synchronously by using the `call` method, which suspends the calling process until the called process terminates. Each process can be stopped asynchronously by using the `stop` method. There is only one main process, which is started automatically at startup. To simplify the definition of a set of processes, parametrizable process arrays can be defined, sharing the same process body definition that is replicated across different and independent physical processes (see **Def. 3**).

*Def. 3* *ConPro process definition and control using object method applications. Processes are objects that can be parametrized (i.e., synthesis parameters).*

```
open Process;
process pname:
begin
  definitions
  statements
end ⟦ with parameter=value ⟧;

array paname : process[N]
begin
  definitions
  statements
end ⟦ with parameter=value ⟧;

...
     pname.start();
pname.stop();
pname.call();
     paname.[index].start();
...
```

Shared program functions and procedures are implemented with sequential processes, too. Function processes are started by a calling process by using the call method applied to the function process object, which suspends the calling process until the function process has terminated. To serialize concurrent function calling, an additional function call wrapper is attached to a function process (Mutex scheduler).

A process can contain any kind of sequential statement (e.g., data assignments, loops, branches) and parallel data-path assignment statements (bounded blocks), with an optional set of parameters applied to the process block or any block inside a process:

```
open Process;
reg a,b,d : logic[8];
process main
begin
  for i = 1 to 5 do
  begin
    d <- d + i;
  end with unroll;
  begin
    a <- d;
    b <- d;
  end with bind;
end with schedule=basicblock;
```

## IV.3 Data Storage Objects

True bit-scalable data types (TYPE $\beta$) and storage objects (subset $\mathfrak{R}$ of TYPE $\alpha$) are supported. The data width can be chosen in the range $\delta$=1-64 Bit. A data storage object $\mathfrak{R}$ is specified and defined by a cross product of types ($\alpha\times\beta$). Storage objects can be read in expressions and modified in assignments.

**Registers** are single storage elements either used as a shared global object or as a local object inside a process. In the case of a global object, the register provides concurrent read access (not requiring a Mutex guarded scheduler) and exclusive Mutex guarded write access. If there is more than one process trying to write to the register, a Mutex guarded scheduler serializes the write accesses. There are two different schedulers available with static priority and dynamic FIFO scheduling policies.

**Variables** are storage elements inside a memory block either used as a shared global object (the memory block itself) or as a local object inside a process. A variable always provides exclusive Mutex-guarded read and write access to the memory block to which it belongs.

Different variables concerning both data type and data width can be stored in one or different memory blocks, which are mapped to generic RAM blocks. Address management is done automatically during synthesis and is transparent to the programmer. Direct address references or manipulation (aka pointers) are not supported.

The memory data width, always having a physical typeof logic/bit-vector, is scaled to the largest variable stored in memory. To reduce memory data width, variables can be fragmented, which means a variable is scattered over several memory cells.

Different memory blocks can be explicitly created, and variables can be assigned to different blocks.

**Queues** are storage elements with FIFO data transfer order and synchronized inter-process communication objects, too. They are always used as shared global objects. Queues and channels can be read directly in expressions and can be used on the left-hand side of assignments like any other storage object.

**Channels** are similar to queues. But they can be buffered (behaviour like a queue with one cell, depth is 1) or unbuffered (providing only a handshake data transfer).

In **Def. 4** there is a summary of signal, data storage, and object definition statemens.

*Def. 4* *ConPro data storage and signal object definitions and data types with optional parametrization (DT: Data type, width: bit width including sign bit, ⟦..⟧: optional)*

```
reg name : DT ⟦ with parameter=value ⟧;
var name : DT ⟦ in blockname ⟧;
block name;
sig name : DT;
queue name : DT ⟦ with depth=value ⟧;
channel name : DT ⟦ with depth=value ⟧;
TYPE DT = { logic, logic[width], int[width], bool, char }
TYPE OT = { reg, var, sig, queue, channel }
```

Registers, variables, queues, and channels can be defined for product types (arrays and structures) and sum types (enumeration types).

## IV.4 Signals and Components

**Signals** are interconnection elements without a storage model. They provide an interface to external hardware blocks. Signals are used in component structures, too.

Signals can be used directly in expressions like any other storage object. Signals can be read in expressions, and a value can be assigned using the assignment statement, Reading a signal returns the actual value of a signal, but writing to a signal assigns a new value only for the time the assignment statement is executed (active), otherwise, a default value is assigned to the signal. Therefore, there may be only one assignment for a signal.

Signals are non-shared objects, and have no access scheduler. Only one process may assign values to a signal (usually using the `wait for` statement), but many processes may read a signal concurrently. Additionally, signals can be mapped to register outputs using the `map` statement. The definition of a signal object is shown in **Def. 4**.

## IV.5 Arrays and Structure Types

Arrays can be defined for data storage and abstract object types, as shown in **Def. 5**. Arrays can be selected with static (index known at compile time) and dynamic selectors (variable expressions) using the dot bracket notation `.[index]`.

*Def. 5* *ConPro array definition and array element access (size: number of array elements)*

```
array AS: OT[size] of DT ⟦ with parameter=value ⟧;
array AO: object obj[size] ⟦ with parameter=value ⟧;
array AO: process[size] of begin .. end
          ⟦ with parameter=value ⟧;
... AS.[index] ...
... AO.[index].method ...
```

A structure type definition contains only data types (DT), and no object types (OT). There are three different subclasses of structures for different purposes, formally described in **Def. 6**:

**Multi-Type Structure**

The generic structure type binds different named structure elements with different data types to a new user defined data type, the native product type.

**Bit Structure**

This structure sub-class provides a bit-index-name mapping for storage objects. All structure elements have the same data type. The bit-index is either a one bit value or a range of bits. This structure type provides the symbolic/named selection of parts of vector data types (for example, logic vector or integer types) and specifies the bit access of object data.

**Component Structure**

This structure defines hardware component ports, either of a ConPro module top-level port, or of an embedded hardware component (modelled at the hardware level). This structure type can only be used with component object definitions. The component type has the same behaviour as the signal type.

In the case where the object type of a structure instantiation is a register, just *N* independent registers are created. In the case of a variable object type, *N* objects are stored in a RAM block. Arrays from structure types can be created, too. For each structure element, a different array is created. Hardware component port types are defined with structures, too, with the difference that for each structure element, the direction of the signal must be specified. Some care must be taken for the direction: if the component is in lower hierarchical order (an embedded external hardware component), the direction is seen from the external view of the hardware component.

If the component is part of the top-level port interface of a *ConPro* module, it must be seen from the internal view.

*Def. 6* *ConPro structure type definitions and element selection*

```
Define a new structure type with a data type DT specification
for each element:
     type ST: {
       E1: DT;
       E2: DT;
       ...
     };
     reg instance:ST;

Define a new component structure type with data type DT
and signal direction DIR specification for each element.
     type CT: {
       port E1: DIR DT;
       port E2: DIR DT;
       ...
     };

Define a new bit structure type with bit-width specification
for each element.
     type BT: {
       E1: width;
       E2: width;
       ...
     };
Access of type object elements
     ... obj.elem ...
```

The following **Ex. 8** shows the usage of process, object, and data arrays:

*Ex. 8* *ConPro* usage of process, object, and data arrays.

```
open Core;
open Process;
open Barrier;
open Queue;
open System;

object b: barrier;
array d: reg[4] of int[8];
array dataqueue: queue[4] of int[8] with depth=8;
export d;

array p: process[4] of
begin
  reg t : int[8];
  for i = 1 to 5 do
  begin
    b.await ();
    t <- # + 1;
    d.[#] <- t;
    dataqueue.[#] <- t;
    d.[#] <- 0;
  end;
end;

process main:
begin
  b.init ();
  for i = 0 to 3 do
   p.[i].start ();
end;
```

## IV.6 Exceptions and Handling

Like in modern programming languages, the *ConPro* programming model supports the concept of exceptions using signals and exception handling with statements, shown in **Def. 7**.

Exceptions provide the only way to leave a control structure, for example, loops, conditional branches or functions themselves. Exception are abstract signals, which can be raised anywhere and caught within a `try-with` exception handler environment, either within the process/function where the exception was raised, or outside. Thus, exceptions are automatically propagated along a call path of processes and functions using exception state registers if they are not caught within the raising process or function, directly implemented with RT logic.

*Def. 7* *ConPro definition of exceptions and exception handlers*

**Definition of an exception**

```
exception E;
```

**Raising of an exception (inside processes)**

```
... raise E; ...
```

**Catching of raised exceptions**

```
try I with
begin
  when E1: I1;
  when E2: I2;
  ...
end;
```

An typical usage of exception propagation is shown in **Ex. 9**.

*Ex. 9* *Rasing of Conpro exception and exception handling.*

```
open Process; open Core; open Random;

exception DIVBYZERO;
object rnd: random with datawidth=8;
  rnd.seed (1000);

function foo(a:int[8],b:int[8]) return (c:int[8]):
begin
  if b = 0 then raise DIVBYZERO;
  c <- a/b;
end;

process main:
begin
  reg d : int[8];
  reg r : logic[8];
  try
  begin
    for i = 1 to 64 do
    begin
      rnd.read (r);
      d <- foo(d,to_int(r));
    end;
  end
  with
  begin
    when DIVBYZERO: d <- 0;
  end;
end;
```

## IV.7 Functions and Procedures

A function definition consists of a unique function name identifier, the function parameter interface, and the function body with statements. The function body consists of local object definitions (types, data, and some abstract objects) and an instruction sequence. The basic syntax of function definitions and calling of functions is shown in **Def. 8**.

Each function parameter and the set of return value parameters are handled like registers. There is only support for call-by-value semantics. Function argument values are copied to the appropriate parameters prior to the function call, and return values are copied after the function call has completed.Within the function body, all parameters and the (named) return parameter can be used in assignments and expressions like any other register. There is no return statement. The last value assigned to the return parameter is automatically returned.

There are two different types of functions: in-lined and shared. The in-lined function type is handled like a macro definition. Each time a function is applied (called), the function call is replaced by the function body, and all function parameters are replaced by the function arguments (including return value parameters). Shared functions are implemented using a sequential process and the call method with an additional function call wrapper. Each time a shared function is called, the argument values are passed to the function parameters (global registers), and the return value(s), if any, is passed back.

*Def. 8* *ConPro definition and application of functions and procedures*

```
function fname (p1:DT,p2:DT,..) ⟦ return (r1:DT,..) ⟧:
begin
  definitions
  instructions
end;

pname();                procedure call
pname(i,1);
y ← fname1(x);          function call returning a value
{y1,y2} ← fname2(x);    function call returning two
                        values
```

The following short example demonstrates the ease of using functions:

```
open Process;
open Core;

function foo(a:int[8],b:int[8]) return (c:int[8]):
begin
  c <- a+b;
end;
process main:
begin
  reg x,y,z:int[8];
  x <- 5, y <- 6;
  z <- foo(x,y);
end;
```

## IV.8 Modules

A *ConPro* design hierarchy consists of a behavioural module level (Module- B) containing global (shared) objects and processes. A module is mapped to a circuit component with a top-level hardware port interface. Structural modules (Module-S) can be composed of behavioural modules with optional internal interconnect components. Each process (and process level) consists of local (non-shared) objects and a process instruction sequence, specifying the control and data flow. Abstract object types are implemented with abstract object modules (Module-O).

Behavioural modules implement objects and processes. A behavioural module is defined by the source code file itself. Actually, there is only one module hierarchy level, the main module. More source code files can be included using include statements. There are two kinds of behavioural modules: modules embedding objects and processes, and modules providing access to and implementation of abstract data type objects (ADTO). These are primarily inter-process communication and synchronization objects, such as Mutex, semaphore, timer and some communication links. Each ADTO module to be used must be opened using the open statement.

*Def. 9* *ConPro definition of structural modules and structural composition using hardware components and IO ports.*

```
Design toplevel IO port definition:

     type top_io_type:{
       port S1: DIR DT;
     };
     component TOP: top_io_type;
     export TOP;

Define a new structural module with specified name:
module MS

     begin
       import
       component
       structtype
       mapping
     end;

Import of behavioural modules and instantiation of components
(ciruits):

     import MB;
     component C1,C2,..:MB;

Defines and instantiate a port interface of interconnect
component with initial signal mapping:

     type ICT:{ port...};
     component IC:ICT :=
     {
       C1.TOP.S1,
       C1.TOP.S2,...
       C2.TOP.S1,...
     };

Internal interconnect using mapping statements:
          IC.S1 >> IC.S2;
```

Structural modules are used to build System-On-Chip (SoC) circuits from behavioural modules. The definition of the structural composition using IO port structures, component instantiations, and module blocks is shown in Definition **Def. 9**. The structural composition aids in the design of complex SoC circuits by reusing multi-process blocks.

In addition to static port components (with fixed directions) using non-buffered signals, the Ioport module provides dynamic ports with buffer registers (as in microcontrolelrs) supporting partial bit updates, shown in the following example **Ex. 10**:

*Ex. 10 Circuit hardware ports and programmable IO ports (Ioport module)*

```
open Core;
open Process;
open Ioport;

type top_port: {
  port p_io1: inout logic[8];
  port p_io2: inout logic[8];
};
component top: top_port;
export top;

object pt1,pt2: ioport with datawidth=8;
  pt1.interface(top.p_io1);
  pt2.interface(top.p_io2);

process p1:
begin
  reg d1,d2: logic[8];


  pt1.dir(0b00000000);
  pt2.dir(0b11111111);
  pt1.read(d1);
  pt2.read(d2);

  for i = 1 to 8 do
  begin
    d1 <- d1 land 0b00001111;
    d1[4 to 7] <- to_logic(i);
    d2[3] <- 1;
    pt1.write(d1);
    pt1.dir(0b11111000),pt2.write(d2);
    d2[3] <- 0;
    pt1.dir(0b00000000),pt2.write(d2);
    pt1.read (d1);
    while d1[0] = 0 do
      pt1.read (d1);
  end;
end;
```

## IV.9 The ConPro Building Blocks

The overall ConPro design hierarchy and the compositional blocks are shown in **Fig. 6**. Top-level modules are composed of process definitions, global objects (storage, synchronization, communication), and user-defined types, and finally top-level interface ports connecting processes and objects with the outside world. Using components, the structural composition can embed hardware blocks on the top level. Processes are composed of instructions and local storage objects.

To summarize. the ontology of ConPro building blocks consists (in descending order):

1. Top-level module and component with a structural interface specification;
2. Storage, object, component, process, function, and module interface (external hardware components) blocks;
3. External Module Interface (EMI);
4. Sequential instructions in functions and processes, bounded blocks, local storage;
5. Type and exception definitions.

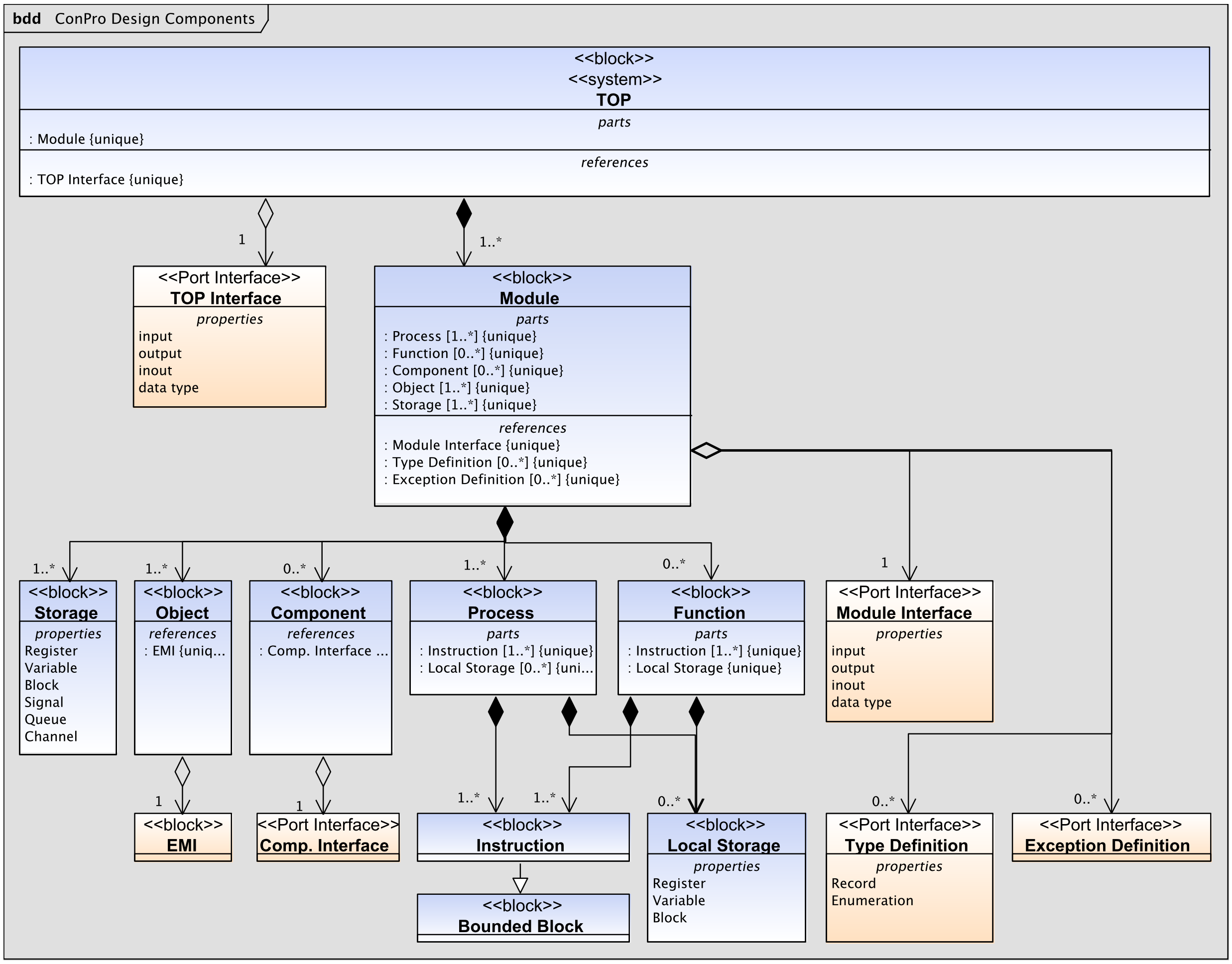

*Fig. 6* *ConPro Design Hierarchy and Compositional Blocks Diagram*

## IV.10 Control and Data Processing Statements

### Data Processing

Data processing is performed by evaluating expressions and storing computation results in registers or variables using assignments. An assignment statement has a target data object (register, variable, or signal), the left-hand side (LHS), and an expression, the right-hand side (RHS), which can be arbitrarily nested and consists of arithmetic, relational, and boolean operations with operands accessing registers, variables, or signals, shown in **Def. 10**.

As in traditional imperative programming models, there is a strict sequential order: first the RHS is evaluated, and then the result is stored, which is commonly ensured by an appropriate hardware model in the data path and can be processed in one control state. Assignments can only be used in process body blocks.

*Def. 10* *ConPro computational statements (assignment and expressions)*

**Assignment**

```
var x: DT; ‖ reg x: DT; ‖ sig x: DT;
x <- expression;
```

Micro-Sequential timing model: *result*=Eval(*expression*) ; Write(*x*, *result*)

**Expressions**

```
expression ::= value | expression operator expression .
operator ::= + - * / % land lor lnot band bor bnot lsl lsr .
```

### Bounded Blocks

Instruction blocks can be used to bind instructions, either part of a control statement like a branch, or used to define a sub-process with a sequential or parallel execution model. Instruction can be parametrized, attaching specific processing or synthesis settings, shown in **Def. 11**. The short form of a bounded block is a comma separated list of assignment instruction statements. All assignments are scheduled in on FSM state and executed in one time step, i.e., variables stored in RAM blocks (typically with two clock cycle access) are replaced by temporary registers. Instruction should not have cross data interdependencies, otherwise the result of the bounded block execution can violate data consistency.

*Def. 11* *ConPro Sequential and Parallel Process Constructor using blocks.*

```
begin  seq. sub-process ⟦ begin ⟧ parallel sub-pro.
  i1;                       i1, i2, i3, ..
  i2;⟦ end ⟧ ⟦ with bind ⟧;
  ..
end with ⟦ parameter=value and .. ⟧;
```

Parameters: { `bind`, `unroll`, `schedule`, `inline`, ..}

### Conditional Branches

Two different conditional branches exist: a mutual branch (if-then-else) based on a boolean condition and multi-case branch matching values with the outcome of an expression evaluation (match-with-when). Mutual branches can be chained (if-then-else-if) offering a prioritized and ordered conditional branching, shown in **Def. 12**.

*Def. 12* *ConPro Branches*

```
if cond then i_cond=true ⟦ else i_cond=false ⟧;

match expr with
begin
  when v1: i_expr=v1;
  when v2,v3, ..: i_expr=v2∨expr=v3..;
  ..
⟦ when others: i_others; ⟧
end;
```

In contrast, the multi-case branch can be parallelized in the control and data paths regarding the evaluation of the matching condition, offering a branch with constant processing time for each case matching.

### Loops

The following loops are offered by the programming model, with the syntax defined below. In addition to common loop statements there are special loops addressing the requirements in parallel process systems with event-based client-server/request-service operational semantic and the handling of hardware logic signals.

**Counting Loop [for-do]**

A counting loop repeats the instruction execution of the loop body with an incrementing or decrementing loop variable. The number of loop iterations can be given by static or dynamic limit values (registers, variables). Loops can be completely unrolled. On top-level, the for-loop can be used to generate an unrolled sequence of top-level instructions, for example, mapping instructions using arrays. Top-level for-loops will always be unrolled, therefore the unroll parameter is unnecessary

**Conditional Loop [while-do]**

The loop body is executed as long as a boolean expression evaluates to the value true.

**Unconditional Loop [always-do]**

The loop body is executed repeatedly without any condition. Used in event-based service processes. An unconditional loop can only be terminated by stopping the process or by raising an exception.

**Constant Delay [wait-for-time]**

The process can be suspended (delayed) exactly for a specified amount of clock cycles or time units.

**Conditional await [wait-for-condition]**

This statement blocks the process until a boolean expression is true. In the case the expression is false and the wait statement is blocked, signal assignments can be applied optionally for this time. The expression generally operates on global signals or registers that may not be guarded.

**Constant clock cycles application [wait-for-time-with]**

For a specified number of clock cycles simple expressions can be evaluated and assigned to signals. A signal on the LHS in an assignment holds a value only for one clock cycle. A signal in this apply statement holds the value for N clock cycles. Outside the application statement the signal must be assigned with a default value. This can be specified optionally in the apply statement

*Def. 13* *ConPro Loops*

```
for x = a to ‖ downto b ⟦ step v_s ⟧ do i_loop ;

while cond do i_loop ;

always do i_loop ;

wait for time;
wait for cond;
wait for time with i_0;
wait for cond with i_0 ⟦ else i_1 ⟧;
```

The following example shows the usage of various loops and branches:

```
...
array links: object uart[NUM_LINKS];
-- Configure UART devices at top-level
for i = 1 to NUM_LINKS do
begin
  links_com.[i-1].interface(DEVLN.links_rx[i-1],DEVLN.links_tx[i-1]);
  links_com.[i-1].baud(115200);
end;

reg d : logic[8];
export d;

process main:
begin
  always do
  begin
    d <- 0;
    while d < 10 do
    begin
      for i = 6 downto 0 do
      begin
        if d[i] = 0 then
          d[i] <- 1;
      end;
    end;
    wait for 1 milisec;
  end;
end;
```

# V Hardware Architecture

The previously introduced extended CCSP model can be directly implemented in hardware and mapped on microchip level **[BOS11A][BOS10A]**.

## V.1 Processes

Each *ConPro* process is implemented at hardware level with a Finite State Machine (FSM) and a Register-Transfer Level (RTL) architecture of the data path, finally creating an interconnected multi-FSM architecture., shown in **Fig. 7**

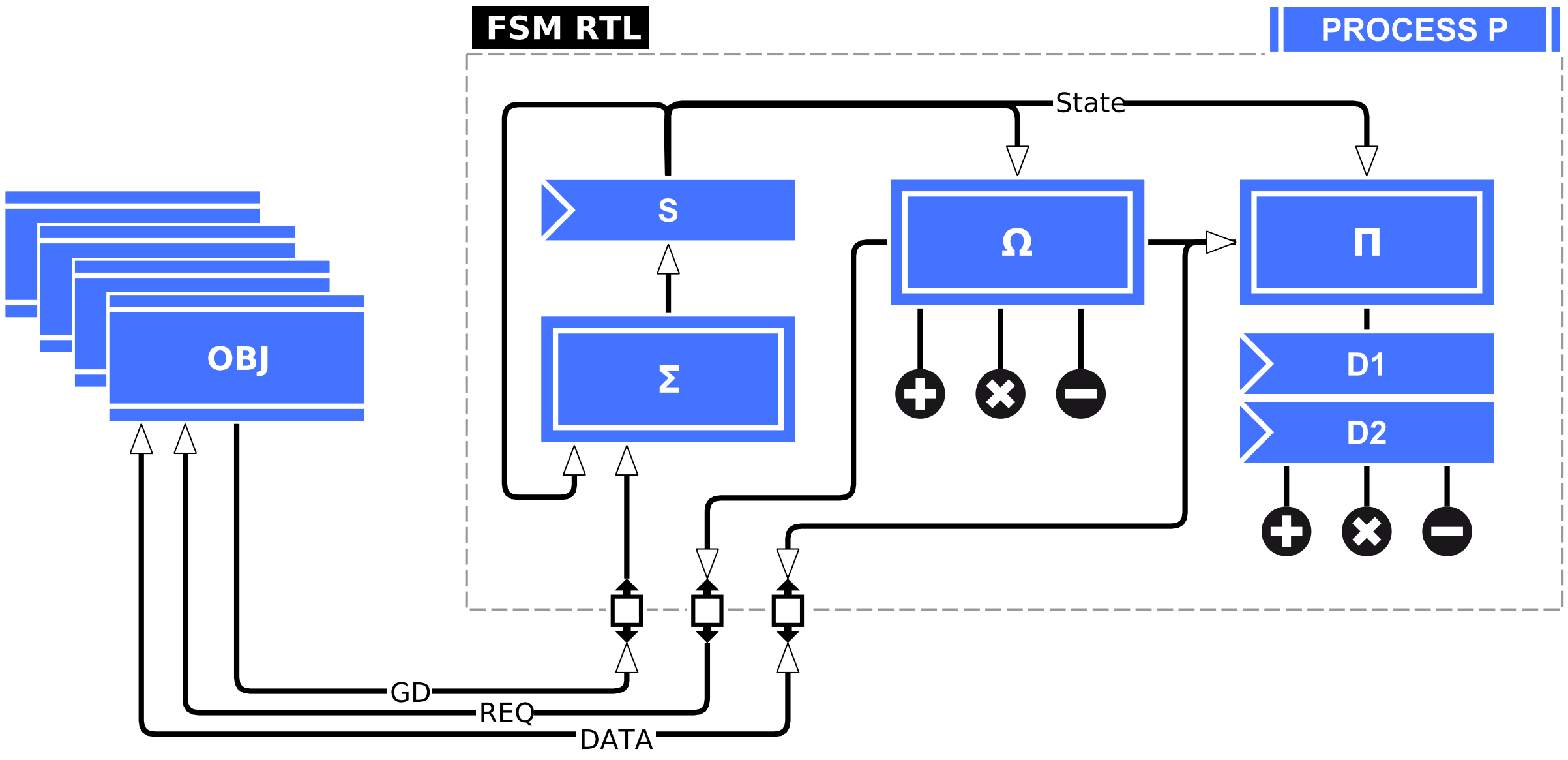

*Fig. 7* *Hardware architecture of a sequential ConPro process and interconnect to global objects (Σ: state transition network, S: control state register, Ω: combinatorial data path, Π: transitional data path)*

The program flow of the instructions of a *ConPro* process is mapped onto the control flow and states of the FSM, with an additional start and end state for each process. Complex instructions are split into multiple states. The data path is divided into a pure combinatorial and a transitional part with registers. The first one accesses local (read only) and global objects (read, write, and control access). Access to global resources is always guarded by the access scheduler (mutual exclusion lock) of the shared object. A request signal (`RE`/`WE` or `CALL`) is activated in the data path. A guard signal `GD` is read and evaluated by the finite state machine transition network. The current state, accessing the object, is maintained until the guard signal changes to a low level (which occurs only once per clock period to avoid race conditions).

Shared program functions and procedures are implemented with sequential processes, too.

## V.2 Modules

All hierarchical modules, processes, objects, and instantiated hardware components are merged into one (flat) SoC design with one top-level port interface, shown in Figure *8*.

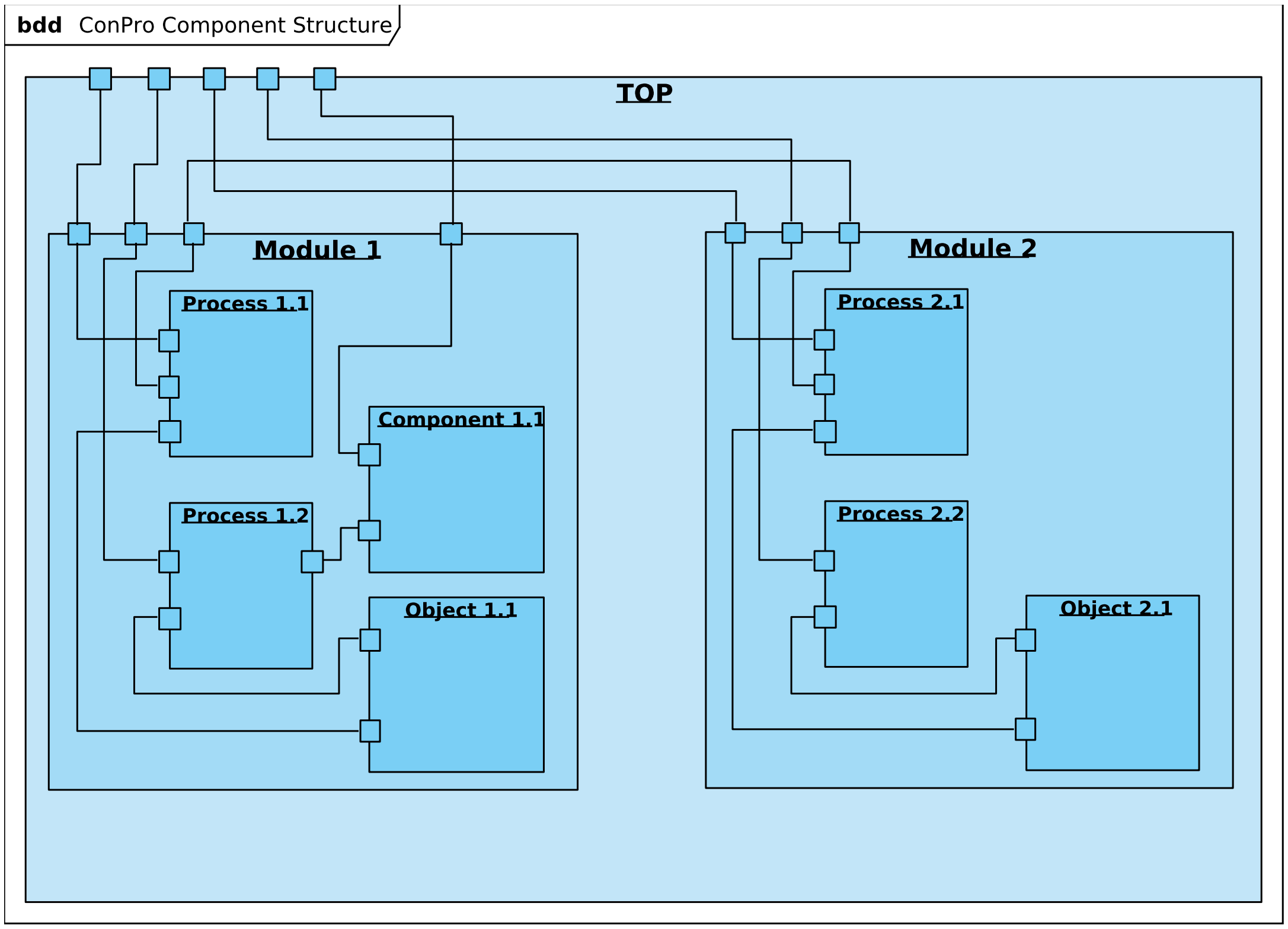

*Fig. 8* *ConPro Hierarchical Block Architecture with Interconnection.*

## V.3 Mutex Scheduler

Access of shared objects must be guarded inherently by a Mutex using a mutual exclusion scheduler. This scheduler is responsible to serialize concurrent access. The scheduler is connected with all processes accessing the resource. A process activates a request (`REQ`), and waits for the release of the guard (`GD`), which unblocks the process signalling the grant of the resource, shown in **Fig. 9**.

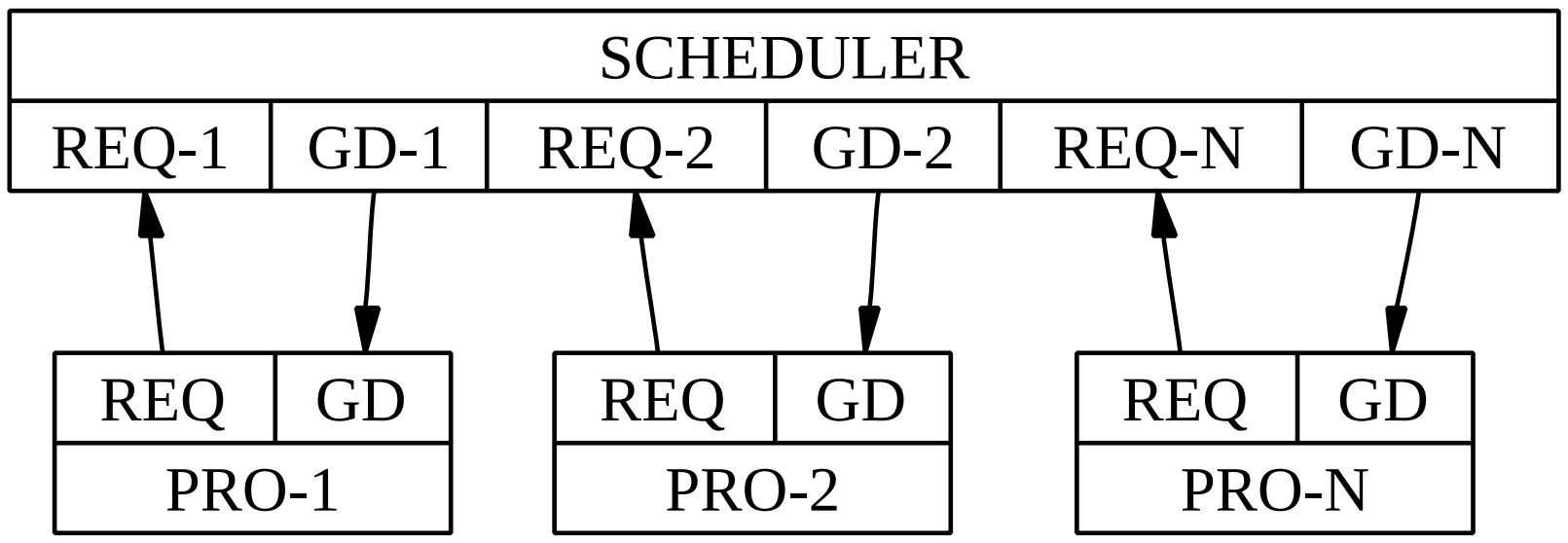

*Fig. 9* *Scheduler Block Architecture and process interconnect*

There are two different schedulers available:

**Static Priority Scheduler [default]**

This is the simplest scheduler and requires the lowest amount of hardware resources. Each process ever accessing the resource gets a unique ordered priority. If there are different processes accessing the resource concurrently, the scheduler always grants access to the process with the highest priority. There is a risk of race conditions using this scheduling strategy. Commonly, the order of processes appearing in the source code deter- mines their priority: the first process gets the highest priority, the last the lowest. A scheduled access requires at least two clock cycles.

**Dynamic FIFO Scheduler**

The dynamic scheduler provides fair scheduling using a process queue. Each process requesting the resource and looses the competition is stored in a FIFO ordered queue. The oldest one in the queue is chosen by the scheduler if the resource is released by the previous owner. The dynamic scheduler avoids race conditions, but requires much more hardware resources.

The algorithms for both schedulers are defined in Definitions **Def. 14** and **Def. 15**.

*Def. 14* *Static Priority Scheduler: From/to process i:{REQ-i,GD-i}, from/to shared resource block:{ACT,ACK}. A process-i request activates REQ-i, and if the resource is not locked, the request is granted to the next process in the if-then-else cascade. If the request is finished, then ACK is activated and releases the locked object and releases GD-i for this respective process indicating that the request is finished.*

```
Loop Do
  ACT ← False;
  ∀ gd ∈ {GD-1,GD-2,..} Do gd ← True;
  If REQ-1 ∧ ¬LOCKED Then
    LOCKED ←True;
    ACT ← True;  Start Service for Process 1
  Else If REQ-2 ∧ ¬LOCKED then
    LOCKED ←True;
    ACT ← True;  Start Service for Process 2
  ...
  Else If ACK ∧ REQ-1 ∧ LOCKED Then
    GD-1 ← False;
    LOCKED ← False;
  Else If ACK ∧ REQ-2 ∧ LOCKED Then
    GD-2 ← False;
    LOCKED ← False;
  ...
```

*Def. 15* *Dynamic Queue Scheduler: From/to process i:{REQ-i,GD-i}, from/to shared resource block:{ACT,ACK}. A process-i request activates REQ-i, and if the resource queue is empty or this process is at head of the queue, the request is granted to the process in the if-then-else cascade. If the request is finished, then ACK is activated and removes the process from the resource queue and releases GD-i for this respective process indicating that the request is finished.*

```
Loop Do
  ACT ← False;
  ∀ gd ∈ {GD-1,GD-2,..} Do gd ← True;

  If REQ-1 ∧ LOCKED=[] ∧ ¬PRO-1-LOCKED Then
    LOCKED ← [PRO-1];
    PRO-1-LOCKED ← True;
    OWNER ← PRO-1;
    ACT ← True;         Start Service for Process 1
  Else If REQ-2 ∧ LOCKED=[] ∧ ¬PRO-2-LOCKED Then
    LOCKED ← [PRO-2];
    PRO-2-LOCKED ← True;
    OWNER ← PRO-2;
    ACT ← True; Start Service for Process 2
  ...
```

```
Else If REQ-1 ∧ LOCKED≠[] ∧ ¬PRO-1-LOCKED Then
  LOCKED ← LOCKED @ [PRO-1]; Append Process 1 to Queue
  PRO-1-LOCKED ← True;
Else If REQ-2 ∧ LOCKED≠[] ∧ ¬PRO-2-LOCKED Then
  LOCKED ← LOCKED @ [PRO-2]; Append Process 2 to Queue
  PRO-2-LOCKED ← True;
...
Else If REQ-1 ∧ Head(LOCKED)=PRO-1 ∧ OWNER≠PRO-1 Then
  ACT ← True;          Start Service for Process 1
  OWNER ← PRO-1;
Else If REQ-2 ∧ Head(LOCKED)=PRO-2 ∧ OWNER≠PRO-2 Then
  ACT ← True;          Start Service for Process 2
  OWNER ← PRO-2;
...
Else If ACK ∧ Head(LOCKED)=PRO-1 Then
  GD-1 ← False;
  PRO-1-LOCKED ← False;
  OWNER ← NONE;
  LOCKED ← Tail(LOCKED);
Else If ACK ∧ Head(LOCKED)=PRO-2 Then
  GD-2 ← False;
  PRO-2-LOCKED ← False;
  OWNER ← NONE;
  LOCKED ← Tail(LOCKED);
...
```

## V.4 Functions

Shared functions are implemented with processes and an additional function scheduler guarding the concurrent access of functions and function parameter. Input and output parameters of functions are implemented with global registers. Only one process can call a function process, at the same time i.e., providing optional input arguments via local registers, starting the function FSM (start state), and waiting for the function process termination (reaching the end state), finally transferring optional return data to local registers..

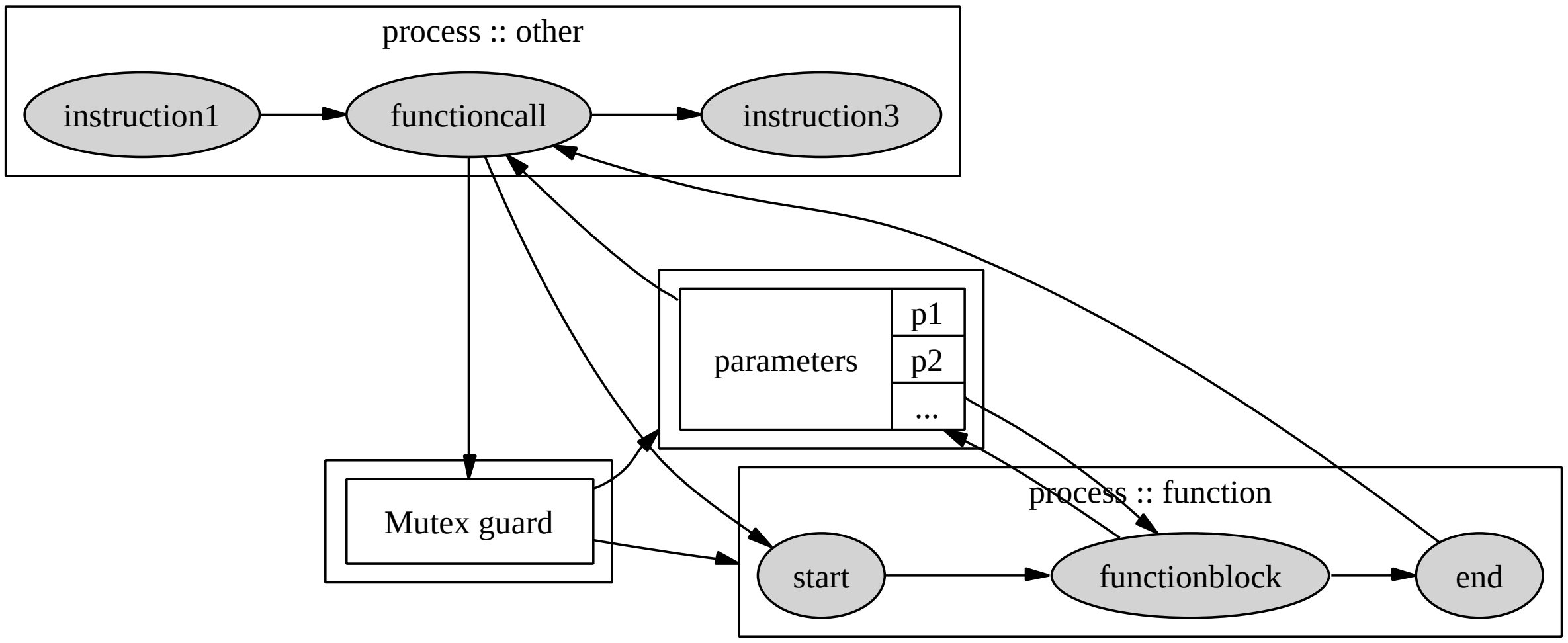

*Fig. 10* *Shared function call using a Mutex scheduler to serialize concurrent function access by different processes.*

The control flow (part of the state diagram of each process) for a function call is shown in **Fig. 10**. Function processes and their Mutex schedulers are connected via hardware signals to the calling processes.

Input signals are the request methods (i.e., function call), input registers (arguments), output signals are the guard signal (granting access to the function) and the output registers (returning values). Function calls typically require only a few additional clock cycles for data transfer and function scheduling.

### V.5 Software Architecture

In addition to the derivation of the proposed hardware architecture model there is the possibility to derive a software model with equal operational and functional behaviour.

Each *ConPro* process is implemented on the software level with a lightweight process (thread), finally resulting in a multi-threaded program with a shared memory model. Global objects are implemented with thread-safe Mutex-guarded global functions. *ConPro* functions can be implemented with thread-safe re-entrant functions without the necessity of a Mutex scheduler.

Different back-ends exist for the *C* or *OCaML* programming languages. Most *ConPro* objects and process statements can be directly implemented in software, except hardware blocks, signals, and statements using signals. The EMI interface can provide software implementations of hardware components.

## VI Case Studies

### VI.1 Dining Philosophers

The Dining Philosophers are a well known problem to demonstrate starvation and deadlocks in parallel and concurrently executing system. **Ex. 11** shows a part of the program code for the implementation of the Dining Philosopher problem. This system consists of five processes simultaneously executing and implementing the actions of the philosophers, i.e., either eating or thinking. For eating they need two forks, i.e., two resources. One resource is shared by two neighbouring philosophers. The philosopher processes are defined using an array construct (line 13). The instructions for each process are processed sequentially. The resource management of the forks is done with semaphores (abstract object type). Each philosopher process tries to allocate two forks, the left- and right-hand side forks, by calling the down method for each semaphore. If a philosopher process succeeds, it calls the (in-lined) function eat, and sets global registers (eating, thinking) simultaneously, as indicated in the program code by using a colon instead of a semicolon (bounded instruction block). An event object ev is used to synchronize the start-up of the group. The philosopher handles event waiting by invoking the await method (line 15). The event is woken up by the main process calling the wakeup method (line 41). All processes begin with the main process (line 40). Processes are treated as abstract objects, too, providing a set of methods controlling the process state.

Objects (like IPC) belong to a module, which has to be opened first (line 1). Each module is defined by a set of EMI implementation files providing all necessary information about the objects of this module (like method declarations, object access, and implementation on the hardware and software level).

*Ex. 11* *Parts of a ConPro source code example: the dining philosopher problem implementation mapped on a multiple processes using semaphores for resource management.*

```
1  open Core; open Process; open Semaphore; open System; open Event;
2  object ev: event;
3  array eating, thinking: reg[5] of logic;
4  export eating,thinking;
5  array fork: object semaphore[5] with
6        Semaphore.depth=8 and Semaphore.scheduler="fifo";
7  function eat(n):
8  begin
9    eating.[n] <- 1,thinking.[n] <- 0;
10   wait for 5;
```

```
  eating.[n] <- 0,thinking.[n] <- 1;
end with inline;
array philosopher: process[5] of
begin
  ev.await ();   synchronize all processes
  if # < 4 then  all processes with array index lower 4
  begin
   always do
   begin
     get left fork then right
     fork.[#].down (); fork.[#+1].down ();
     eat (#);
     fork.[#].up (); fork.[#+1].up ();
   end;
  end
  else
  begin
   always do
   begin
     -- get right fork then left
     fork.[4].down (); fork.[0].down ();
     eat (#);
     fork.[4].up (); fork.[0].up ();
   end;
  end;
end;
process main:
begin
  for i = 0 to 4 do
    philosopher.[i].start ();
  ev.wakeup ();   start the game ...
end with schedule="basicblock";
```

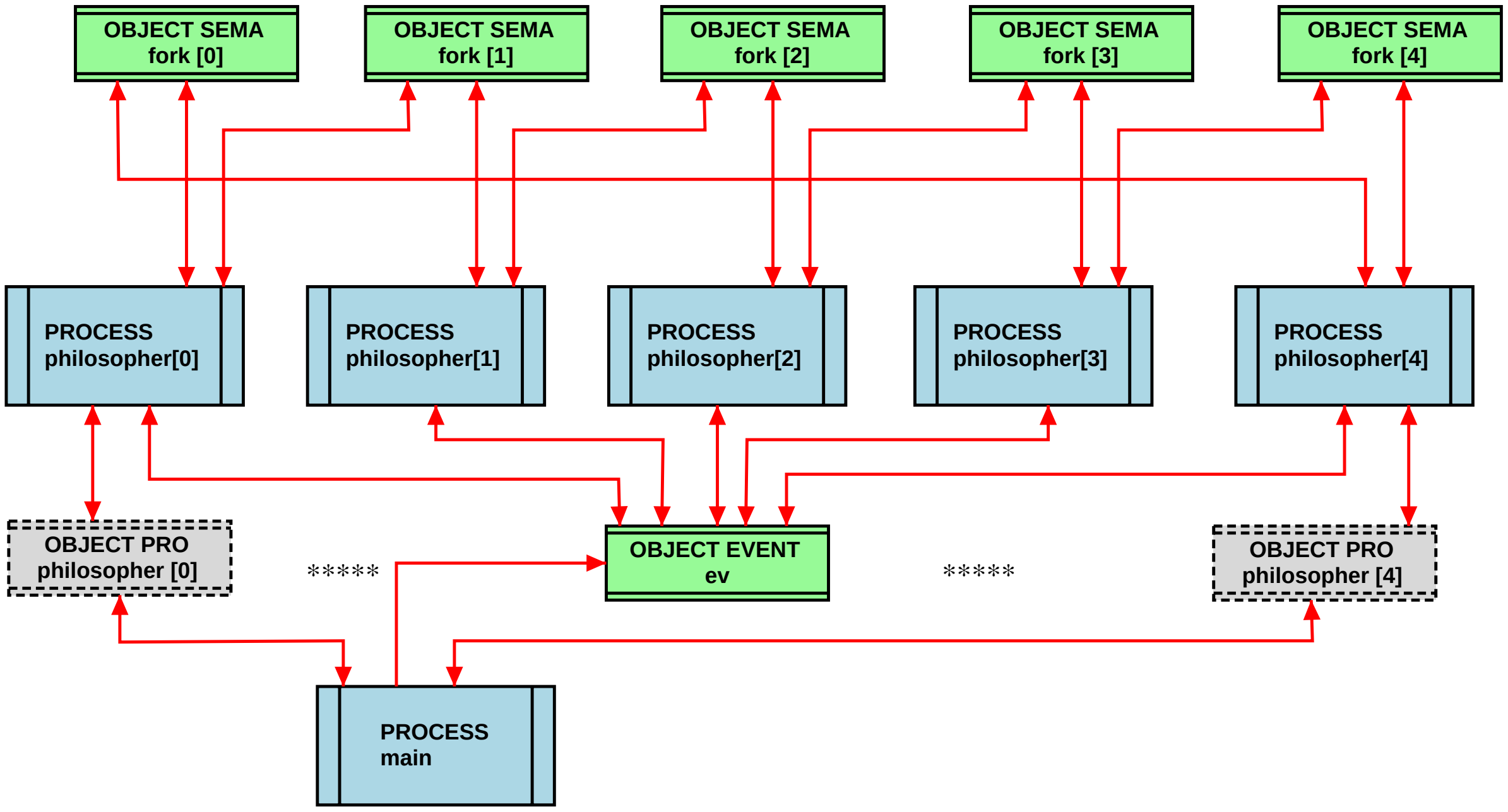

*Fig. 11 Process and inter-process-communication architecture of the dining philosopher problem implementation from Example***Ex. 11**.

The structural CCSP model is shown in **Fig. 11**. Processes are objects and execution units, too. Each global shared object is guarded by a Mutex scheduler that resolves concurrent access by serialization.

This artificial circuit can be directly implemented in an FPGA. The monitor registers *eating* ans *thinking* are exported to the outside ports of the circuit and their signals can be observed externally. The event object *ev* is used to synchronize the start of all philosopher processes (exact synchronous clock event). Assuming that the control-path FSMs of the processes are clocked synchronously, the dynamic behaviour of the circuit should be always equal. The default scheduler behaviour of an ADTO (semaphore) is a prioritized mutex scheduler using a prioritized multiplexer, i.e., the process number determines the winner of a concurrent access. The scheduling order is commonly given by the reverse order of the processes accessing the respective object resource, i.e., the last process is the first that can win the competition.

Since the philosopher processes access the forks in a ring, only two processes are in competition for one fork. Process #0 acquires semaphores 0 and 1, process #1 acquires semaphores 1 and 2, and so on. With the configuration given above, semaphore #0 will result in a process priority order (p#4 > p#0), semaphore #1 has (#p1 > p#0), semaphore #2 has (p#2 > p#1), and so on. So always the "left" philosopher will win the competition (last first process number). This will result always in a deadlock situation in the first round.

## VI.2 Adaptive Communication Protocol Router SLIP

In this section a complete implementation of the adaptive routing SLIP protocol stack is shown here **[BOS11A]**. SLIP is scalable with respect to the network size (address size class (ASC), ranging from 4 to 16 bits), maximal data payload (data size class (DSC), ranging from 4 to 16 bits), and the network topology dimension size (address dimension class (ADC), ranging from 1 to 4 bits). Network nodes are connected using (serial) point-to-point links, and they are arranged along different metric axes of different geometrical dimensions: a one-dimensional network (ADC=1) implements chains and rings, a two-dimensional network (ADC=2) can implement mesh grids, a three-dimensional network (ADC=3) can implement cubes, and so on. Both incomplete (missing links) and irregular networks (with missing nodes and links) are supported for each dimension class, as shown in **Fig. 12**.

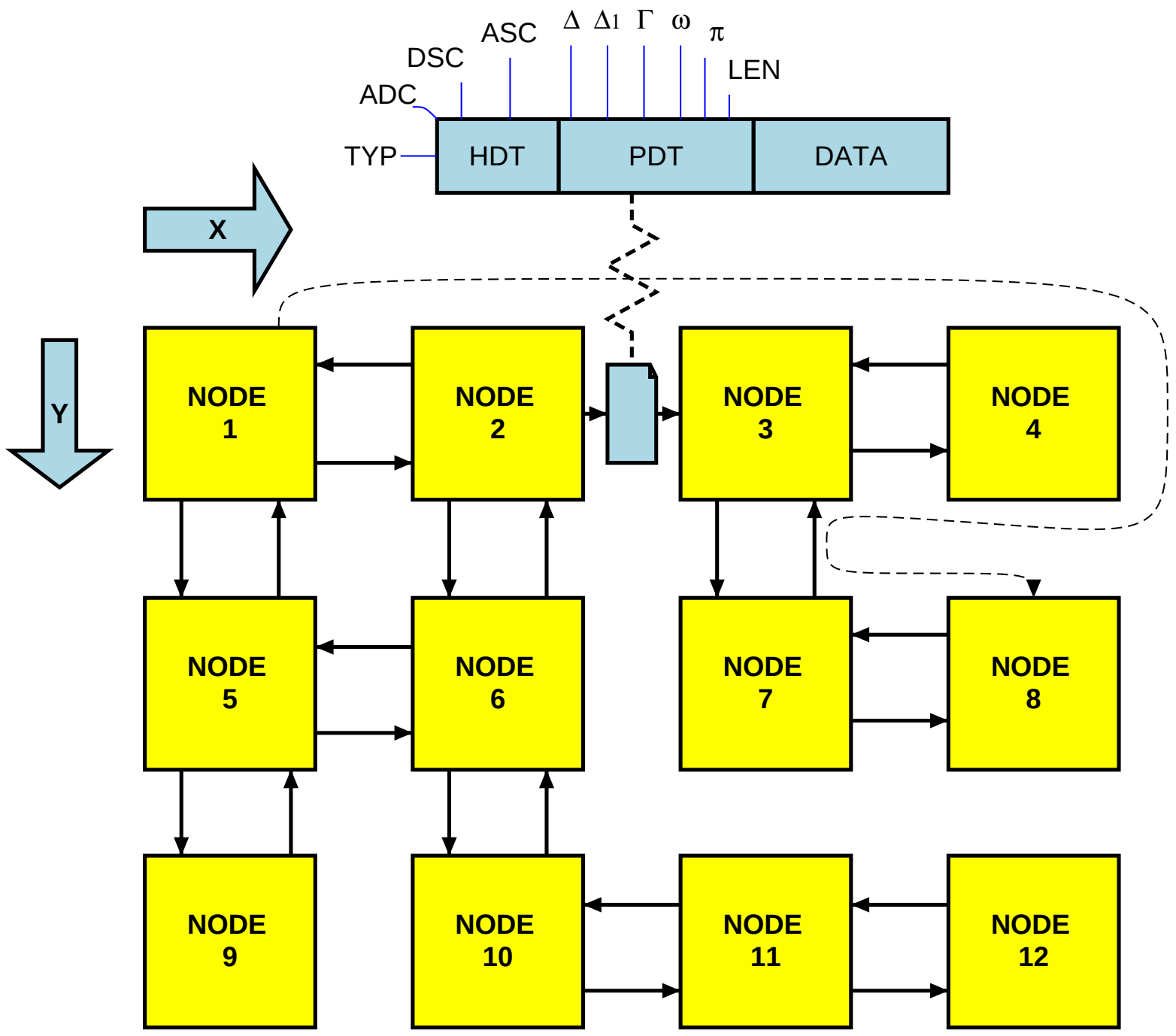

*Fig. 12* *Message-based communication in two-dimensional networks using delta-distance vector routing. Networks with incomplete (missing links) and irregular (missing nodes) topologies are supported by using smart routing.*

One major challenge in message-based communication is the routing and, thus, the unique addressing of nodes. Absolute and unique addressing of nodes in a high-density sensor network, e.g., like in material-integrated damage diagnostic systems, is not suitable. An alternative routing strategy is relative delta-distance routing used by SLIP. A delta-distance vector Δ specifies the way from the source to a destination node, counting the number of node hops for each dimension.

A message packet contains a header descriptor specifying the type of the packet and the scalable parameters ASC, DSC, and ADC, shown in **Tab. 4**. The network address dimension ADC and the size class ASC reflect the network topology, the data size class DSC represents the data payload. There are two main types of messages: requests and replies.

*Tab. 4 SLIP message format (HDT: header descriptor, PDT: packet descriptor)*

| Entry | Size [bits] | Description |
|---|---|---|
| HDT:ADC | 2 | Address Dimension Class |
| HDT:ASC | 2 | Address Size Class |
| HDT:DSC | 2 | Data Size Class |
| HDT:TYP | 2 | Message type = {Request, Reply, Alive, Acknowledge} |
| PDT:Δ | Num(ADC)*Bits(ASC) | Actual delta vector |
| PDT:Δ^0^ | Num(ADC)*Bits(ASC) | Original delta vector |
| PDT:Γ | 2*Num(ADC) | Backward propagation vector |
| PDT:ω | Bits(ADC) | Preferred routing direction |
| PDT:π | Bits(ASC) | Application layer port |
| PDT:LEN | Bits(DSC) | Length of packet |
| DATA | LEN*Bits(DSC) | Payload data |

A packet descriptor follows the header descriptor, containing: the actual delta-vector Δ, the original delta-vector $\Delta^0$, a preferred routing direction ω, an application layer port π, a backward-propagation vector Γ, and the length of the following data part. The total bit length of the packet header depends on the {`ASC`, `DSC`, `ADC`} scalable parameter tuple settings, which optimize application-specific overhead and energy efficiency (spatial & temporal). Each time a packet is forwarded (routed) in some direction, the delta-vector is decreased (magnitude) in the respective dimension entry.

Routing in x-direction, for example, yields: $\Delta_1 = \Delta_{1,-1}$. A message has reached the destination iff Δ=0 and can be delivered to the application port π. There are different smart routing rules, applied in the order shown below, until the packet can be routed (or discarded), as shown in **Def. 16**. First, the normal XY routing is tried, where the packet is routed in each direction one after another with the goal of minimizing the delta count in each particular direction. If this is not possible (due to a lack of connectivity), the packet is sent in the opposite direction, as indicated in the message packet descriptor's gamma entry Γ. Backward routing is used to avoid small area traps, while opposite routing is used to avoid large area traps or to return the packet to the source node (packet not deliverable). The routing decision is made based on the actual message entries {Δ, Γ, ω}, and adaptive routing is achieved by reflecting the actual network topology and the path the message has already travelled, including back-end traps, resulting in alternative paths by selecting different routing directions.

*Def. 16 Smart Routing Protocol SLIP (simplified)*

```
M: Message(Δ,Δ 0 ,Γ,ω,π,Len,Data)
PRO smart_route(M):
  IF Δ=0 THEN DELIVER(M,π) ELSE
  TRY route_normal(M) ELSE
  TRY route_opposite(M) ELSE
  TRY route_backward(M) ELSE DISCARD(M);
PRO route_normal(M):
  FORONE δ i ? Δ TRY minimize δ i :
  route(Δ,M) WITH δ i :=( δ i +1)|δ i<0 ?( δ i −1)|δ i>0 ;
```

```
PRO route_opposite(M):
  FOREONE δ i ? Δ TRY minimize δ i :
  route(Δ,M) WITH δ i :=( δ i –1)|δ i<0 ?( δ i +1)|δ i>0 ;
PRO route_backward(M):
  SEND M (received from direction δ i )
  back to direction -δ i WITH Γ i =-δ i /|δ i |;
```

SLIP implements smart routing of messages with Δ-addressing of nodes arranged in an n-dimensional network space (line, mesh, cube) using peer-to.-peer links. The network can be heterogeneous regarding node size, computation power, and memory. The communication protocol is scalable in terms of network topology and size. A node in a network is a service endpoint and a router, too, which must be implemented on each network node. But the router requires only information about connectivity to direct neighbouring nodes.

To summarize, the routing information is always kept in the packet, consisting of:

1. A header descriptor (*HDT*) specifying the address size class *ASC*, the address dimension class *ADC* (for example 2 is a two-dimensional mesh-grid);
2. A packet descriptor (*PDT*) with routing and path information, and finally the data part.

The SLIP protocol implementation must deal with the variable format of a message. SLIP was designed for low-resource System-on-Chip implementations using ASIC and FPGA target technologies, but a software version was also required. To minimize the resource requirement, the hardware implementation will usually be limited in the dimensional and address spaces, supporting only a subset of the message space, e.g. `ADC`=2, `ASC`=8 (that means δ=-128,..,127).

A node should handle several serial link connections and incoming packets concurrently, thus the protocol stack is a massive parallel system, and was implemented with the *ConPro* behavioural multiprocess model.

The programming model implementation with partitioning of the protocol stack in multiple processes executing concurrently and communicating using queues is shown in **Fig. 13**.

Each link is serviced by two processes: a message decoder for incoming and an encoder for outgoing messages. A packet processor `pkt_process` applies a set of smart routing computation functions (`route_normal`, `route_opposite`, `route_backward`, applied in the given order until routing is possible), finding the best routing direction. Communication between processes is implemented with queues. There are three packet pools holding HDT, PDT and data parts of packets. They are implemented with arrays. The packet processor can be replicated to speed up processing of packets.

The SLIP implementation is partitioned into a receiver state machine and a sender encoder. The receiver state machine parses an incoming byte stream sequentially and incremental, shown in **Ex. 12**.

*Ex. 12 SLIP receiver state machine (simplified)*

```
process pkt_process:
begin
  --
  -- Process incoming data stream and
  -- parse packet header
  --
  type states: {
    S_rx_hdt;
    S_rx_pdt_d1;
    S_rx_pdt_d2;
    S_rx_pdt_d3;
```

```
    S_rx_pdt_dp;
    S_rx_pdt_gam;
    S_rx_pdt_len;
    S_rx_data;
    S_rx_deliver;
    S_discard;
    S_alive;
  };

  reg state: states;
  state <- S_rx_hdt;
  always do
  begin
    match state with
    begin
     when S_rx_hdt:
       {d,tmo} <- link_read();
       ind <- get_rx_pkt(d);   -- Allocate packet structure form pool
       typ <- d[6 to 7],asc <- d[4 to 5],dsc <- d[2 to 3],adc <- d[0 to 1];
       if adc <> ADC_CODE or
         dsc <> DSC_CODE or
         asc <> ASC_CODE then
         state <- S_discard
       else if typ = HDT_TYP_ALV or typ = HDT_TYP_ACK then
         state <- S_alive
       else
         state <- S_rx_pdt_d1;
     when S_rx_pdt_d1:
       if ASC = 4 then
       begin
        {d,tmo} <- link_read();
        if tmo = true then raise Rx_error;
        pkt_pool_pdt.[ind].pdt_d1 <- to_int(d[0 to 3]);
        pkt_pool_pdt.[ind].pdt_d1_hop <- to_int(d[4 to 7]);
        if ADC > 1 then
          state <- S_rx_pdt_d2
        else
          state <- S_rx_pdt_dp;
       end;
       if ASC = 8 then
       begin
         for n = 1 to 2 do ...
       ...
     when S_rx_data:
       {d,tmo} <- link_read();
       pkt_pool_data.[off] <- d;
       off <- off + 1, count <- count + 1;
       if count = len then
         state <- S_rx_deliver;
    ...
    end;
  end;
end;
```

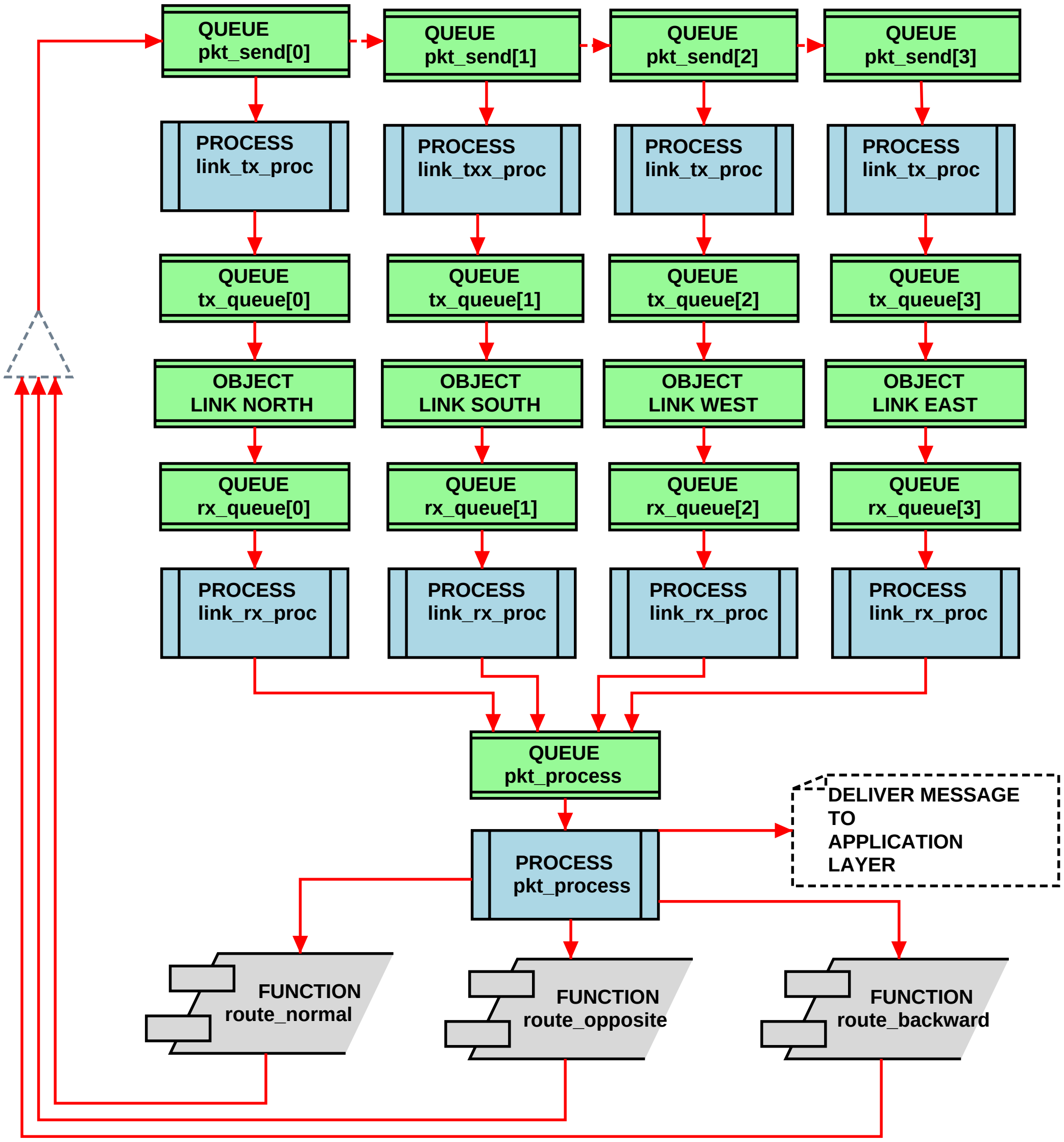

*Fig. 13* *Process and inter-process-communication architecture of the SLIP protocol stack.*

A test setup consisting of the routing processing part of *SLIP* was implemented A.) in hardware (RTL-SoC, gate-level synthesis with Mentor Graphics Leonardo Spectrum and SXLIB standard cell library), and B.) in software (*SunOS*, *SunPro C* compiler). A packet with ADC=2, Δ=(2,3) and a link setup of the node L=(-y,-x) is received on the second link (-x) [L01] and is processed first by the `route_normal` rule (would require connected +x /+y links) [L03], and finally by the `route_opposite` rule [L04] forwarding the modified packet to the link_0 process [LA0].

Tables **Tab. 5** and **Tab. 7** show synthesis and simulation results for the hardware (HW) and the operational equivalent software (SW) implementation. They show low resource demands and latency. Different checkpoints Lxx indicate the progress of packet processing. Figures in brackets give the latency progress relative to the previous checkpoint.

*Tab. 5* *Comparison of resources required for the HW implementation of the routing part of SLIP implemented with a packet pool: (1) variable array, (2) register array. ASIC synthesis was performed with Leonardo Spectrum software and SXLIB standard cell library*

| Resource | Variable1 | Register2 |
|---|---|---|
| Registers [FF] | 767 | 587 |
| Area [gates] | 12475 | 10758 |
| Path delay [ns] | 18 | 16 |
| Synthesized Source CP ⟶ VHDL | 1109 ⟶ 9200 lines | 1109 ⟶ 7900 lines |

Two different storage resource models are compared: variable arrays with RAM blocks and register arrays. Surprisingly, the register storage model is more resource-efficient than the RAM storage model. But this fact holds only for ASIC technologies, and not for FPGA technologies with predefined functional units and already contained on-chip block RAM resources. The register storage model leads to lower computational latency of the parallel packet processing due to the CREW access behaviour. The full design implementation with a *Xilinx* Spartan III (1000k eq. gates) FPGA and the ASIC synthesis is compared in **Tab. 6**.

*Tab. 6* *Comparison of resources required for the full HW implementation of SLIP including simple application layer: (1) FPGA synthesis was performed with Xilinx ISE, (2): ASIC was performed with Leonardo Spectrum software and SXLIB standard cell library*

| Resource | FPGA1 | ASIC2 |
|---|---|---|
| Registers [FF] | 2925 | 15000 |
| LUT [#] | 11261/15360 | - |
| Area [gates] | - | 244 k gates ≈ $2.5mm^2$ \| 0.18µm |
| *Conpro* Source | 4000 lines, 34 processes, 30 shared objects (16 queues, 2 timers) | |
| Synthesized VHDL | 32000 lines | |

*Tab. 7* *Simulation results of the HW and SW implementation of the routing part of SLIP. HW: packet pool: (1) variable array, (2) register array, clock cycles. SW: SunPro CC, SunOS, USIII, CPU machine operations*

| Checkpoint | Clock Cycles1 | Clock Cycles2 | Machine Operations |
|---|---|---|---|
| L01 | 104 | 102 | 60000 |
| L03 | 113 (δ=9) | 107 (δ=5) | 60019 (δ=19) |
| L04 | 187 (δ=74) | 148 (δ=41) | 60796 (δ=777) |

| Checkpoint | Clock Cycles1 | Clock Cycles2 | Machine Operations |
|---|---|---|---|
| LA0 | 235 (δ=48) | 184 (δ=36) | 62305 (δ=1509) |

## VI.3 Parallel Code Morphing VM uFORTH

In contrast to common program-controlled processors executing static programs that exchange data via communication, μFORTH bases on mobile code and code morphing, providing adaptive machine code with embedded data. The program itself can modify the code (and embedded data), which can be sent to other processors.

This stack-based runtime environment is specifically designed for the implementation of mobile agents by using dynamic code morphing under the constraints of low-power consumption and high component miniaturization (but not limited to). It uses a modified and extended version of the FORTH programming language for mobile and adaptive code. FORTH is a stack-based interpreted language whose source code is extremely compact (most operations are zero-operand machine instructions). Furthermore, FORTH is extensible. Code and users can define new language constructs (called words, zero-operand functions) at runtime.

A FORTH program contains built-in core instructions directly executed by the FORTH processing unit and user-defined high-level word and object definitions, which are added to and looked up from a dictionary table. This dictionary plays a central role in the implementation of distributed systems and interacting nodes in sensor networks. Words can be added, updated, and removed (forgotten), controlled by the FORTH program itself or the platform. User-defined words are composed of a sequence of words.

The principal system architecture of one FORTH processing unit (PU) part of the node run-time environment is shown in **Fig. 14**.

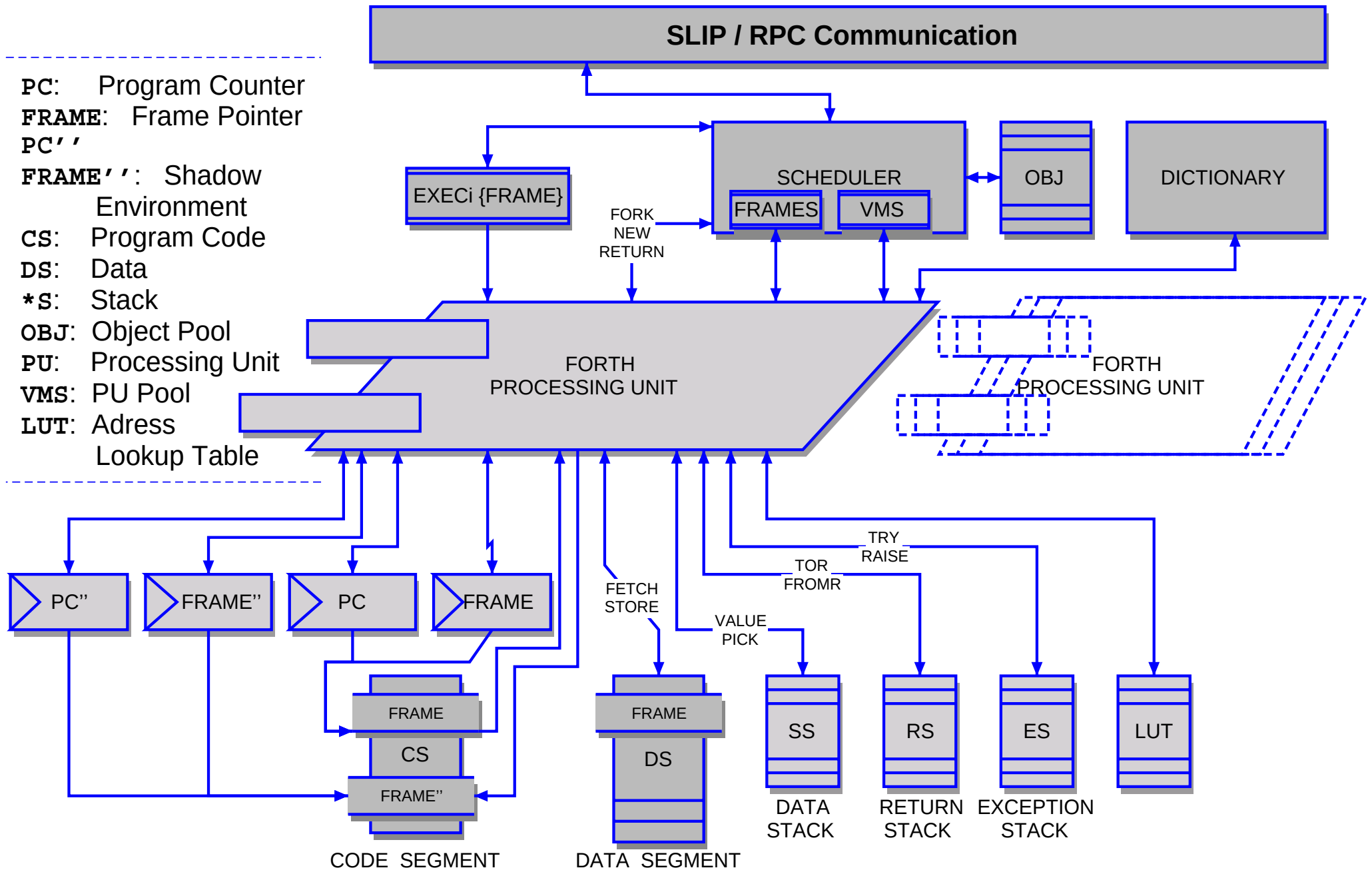

*Fig. 14* *Run-time architecture consisting of μFORTH data processing units, shared memory and objects, dictionary, scheduler, and the SLIP communication router.*

A complete run-time unit consists of a communication system with a smart routing protocol stack connecting processors, one or more FORTH processing units with a code morphing engine, resource management, code relocation, and dictionary management, and a scheduler managing program execution and distribution, which are normally part of an operating system that does not exist here.

A FORTH processing unit initially waits for a frame (a FORTH program) to be executed. During program execution, the FORTH processing unit interacts with the scheduler to perform program forking, frame propagation, program termination, object creation (allocation), and object modification. The set of objects consists of the Interprocess-Communication objects (IPC: mutex, semaphore, event, timer) and frames. There are private and public (node-visible) variables and arrays. All program frames have access to public variables by looking up references stored in the dictionary. Program word, memory variable, and object relocation are carried out by using a frame-bound lookup table LUT.

The scheduler is the bridge between a set of locally parallel-executing FORTH processing units, and the communication system, a remote procedure call (RPC) interface layered above SLIP, a fault-tolerant message-based communication system used to transfer messages (containing code) between nodes using smart delta-distance-vector routing with the previously introduced SLIP router.

The simple FORTH instruction format is an appropriate starting point for code morphing, i.e., the ability of a program to modify itself or make a modified copy, mostly as a result of a previously performed computation. Calculation results and a subset of the processing state can be stored directly in the program code, which changes the program's behaviour. The standard FORTH core instruction set was extended and adapted for the implementation of agent migration in mesh networks with a two-dimensional grid topology. A FORTH program is contained in a contiguous memory fragment called a code frame.

The simple FORTH instruction format (text and binary code) is an appropriate starting point for code morphing, i.e., the ability of a program to modify itself or make a modified copy, mostly as a result of a previously performed computation. Calculation results and a subset of the processing state can be stored directly in the program code, which finally changes the program's behaviour. The standard FORTH core instruction set was extended and adapted for the implementation of agent migration in mesh networks with two-dimensional grid topologies. In our system, a FORTH program is contained in a contiguous memory fragments called frames.

A frame can be transferred to and executed on remote nodes and processing units. **Tab. 8** gives a summary of the new words provided for code morphing. These instructions can be used to modify the program's behaviour and enable the preservation of the current program execution state. In our system, a FORTH program is contained in a contiguous memory fragment, i.e., a code frame. A frame can be transferred to and executed on remote nodes and processing units. Modification of the program code is always performed in a shadow frame environment, which can be identical with the execution frame. This is the default case used for in-place code modification. One or more different frames can be allocated and used for out-of-place modification, which is required if the execution frame is used beyond code morphing. All code morphing instructions operate on the shadow frame. Both the execution and the shadow frames have their own code pointers.

*Tab. 8* µ*FORTH extensions for code frame management and modification*

| Word | Description |
|---|---|
| c! ( frame -- )<br>RESET c! | SETC: Sets frame of shadow environment for code morphing; `RESET` sets code pointer of shadow frame to the beginning of shadow frame |
| >>c ( m1 m2 -- ) | COPYC: Switches to morphing state: Transfers code from program frame between two markers *m*~1~ and *m*~2~ into a shadow frame (including markers) |
| >c ( -- ) | TOC: Copies next word from program frame into shadow frame |
| s>c ( n -- ) | STOC: Pops *n* data value(s) from stack and stores values as word literals in shadow frame |
| <m> ( -- ) | MARKER: Sets a marker position anywhere in a program frame |
| <m>@ ( -- marker ) | GETMARKER: Returns a marker (maps symbolic names to unique numbers) |
| <m>! ( -- ) | SETMARKER: Sets shadow code pointer after marker in shadow frame. Marker is searched in shadow frame, thus either in-place of execution frame or in a new created/copied shadow frame (containing already code and marker). Can be used to edit a partial range of shadow frame code using STOC and TOC instructions |

*Tab. 9* µ*FORTH extensions providing 1. dictionary modification and object creation, and 2. multi-processing support and frame distribution.*

| Word | Description |
|---|---|
| variable x<br>array [n,m] x<br>variable* x<br>aerray* [n,m] x | Creates a new variable or array and allocates memory. The first two definitions create public objects and they are added to the dictionary. The star definitions create private objects. |
| object MUTEX x<br>object FRAME f | Creates (allocates) a new IPC or frame object. The object is added to the dictionary. Other supported IPC object types: SEMA, EVENT, TIMER. |
| import variable x<br>import object x | Imports a variable or object from the dictionary. If not found, then the program execution terminates (return status 0) |
| fork ( dx dy -- ) | Sends content of shadow frame for execution to node relative to actual node. If *dx*=0 and *dy*=0, the shadow frame is executed locally and concurrently on a different FORTH processing unit. The fork instruction returns the frame sequence or processing unit number |
| join (id -- ) | Waits for termination of a forked frame or the reception and execution of a reply program frame. |
| status return | Finishes execution of a program. If status is zero, no reply is generated. If status is equal to -1, an empty reply is generated. Finally, if status is equal 1, the content of the shadow frame is sent back to the original sender of the execution frame |

All architecture parts of the multiprocessor-FORTH node, including SLIP communication, μFORTH processing units, scheduler, dictionary, and relocation support, are mapped entirely to hardware at the multi-RT level and a single SoC design using the ConPro compiler. The resource demand depends on the choice of design parameters and is in the range of 1M–3M equivalent gates (in terms of FPGA architectures). The entire design is partitioned into 43 concurrently executed sequential processes, communicating by using 24 queues, 13 mutexes, 8 semaphores, 52 RAM blocks, 59 shared registers, and 11 timers.

Some implementations details are following. The first simplified code snippet in **Ex. 13** shows the implementation of the word dictionary, basically consisting of the function *lookup* (lines 44-69) and *def* (lines 17-42, basically using the lookup algorithm). The dictionary is store in a RAM block (var storage class), name strings are implemented with compact logic bit vectors.

*Ex. 13 μFORTh word dictionary (simplified, max. word name length is 8 charcters == 64 bit)*

```
1  type dictionary_row : {
2    dict_flag: logic[4];
3    -- Word name, 8 characters, packed format
4    dict_name: logic[64];
5    -- Frame number (required for frame LUT)
6    dict_frame: int[FRAMES_WIDTH];
7    -- PC start address relative to frame base
8    dict_pc: int[FRAME_SIZE_WIDTH];
9    -- Length of word definition
10   dict_len: int[FRAME_SIZE_WIDTH];
11 };
12 ...
13 array dict: var[DICT_SIZE] of dictionary_row;
14 exception Found;
15
16 -- Add new word to dictionary
17 function def(name: logic[64], kind: row_flags, frame:int[FRAMES_WIDTH],
18              len:int[FRAME_SIZE_WIDTH], off:int[FRAME_SIZE_WIDTH]):
19 begin
20   try
21   begin
22     for i = 0 to DICT_SIZE-1 do
23     begin
24       if dict.[i].flag <> ROW_EMPTY then
25         -- Same as lookup, search for existing word entry --
26         ..
27       -- not found, new entry
28       dict.[cfree].name <- name;
29       dict.[cfree].frame <- frame;
30       dict.[cfree].off <- off;
31       dict.[cfree].len <- len;
32   with
33   begin
34     when Found:
35       begin
36         -- only update entry
37         dict.[dind].frame <- frame;
38         dict.[dind].off <- off;
39         dict.[dind].len <- len;
40       end;
```

```
  end;
end;

function lookup(name: logic[64], forget: bool)
  return (found:bool, frame:int[FRAMES_WIDTH], pc:int[FRAME_SIZE_WIDTH]):
begin
    reg dind: int[DICT_SIZE_WIDTH];
    try
    begin
      for i = 0 to DICT_SIZE-1 do
      begin
        dind<-i;
        if dict.[i].flag <> ROW_EMPTY then
        begin
         if name = dict.[i].name then raise Found;
        end;
      end;
      found <- false;
    end
    with
    begin
      when Found:
      begin
        frame <- dict.[dind].frame;
        pc <- dict.[dind].off;
        found <- true;
      end;
    end;
  end;
```

The next code snippet in **Ex. 14** shows the main execution loop. Since the stacks are private storage only accessed by the VM bytecode interpreter, they are defined inside the *vmexec* process. Each stack is split into single registers (top of stack) and an var array holding deeper elements. Writing to registers require only one clock cycle, reading can be done immediately, in contrast, to variables stored in RAM blocks, typically requiring two clock cycles for read and write access. Multiple registers can be accessed concurrently, whereas RAM cells can only be accessed sequentially.

*Ex. 14* µ*FORTH main execution loop*

```
...
type uforth_commands : {
  ------------------------------------------
  -- Data Stack CMD_STACK
  ------------------------------------------
  DUP;      -- (S)
  QDUP;
  ------------------------------------------
  -- Arithmetic Operations CMD_STACK
  ------------------------------------------
  ADD;      -- (S)
  SUB;      -- (S)
  -- ------------------------------------------
  -- Frame Processing CMD_CONTROL
  -- ------------------------------------------
```

```
  FORK;
..
};
process vmexec:
begin
  always do
  begin
    while ret = false do
    begin
      -- Get next instruction
      word <- cs.[pc];
      -- Big bounded block command in one FSM state and clock cycle!
      cmd <- word[0 to 11],data <- to_int(word[0 to 11]),
      wd <- word[12 to 15] lor morphing,
      sm <- BAL, rm <- BAL,
      t0'<-t0,t1'<-t1,t2'<-t2,r0'<-r0,r1'<-r1,r2'<-r2,
      tn_p_s1 <- (tn_p-1) land SS_SIZE_MASK,
      rn_p_s1 <- (rn_p-1) land RS_SIZE_MASK,
      tn_next_a1 <- (tn_next+1) land SS_SIZE_MASK,
      rn_next_a1 <- (rn_next+1) land RS_SIZE_MASK,
      pc <- pc + 1;
      match wd with
      begin
        -- VAL => STACK
        when VAL:
        begin
          t0 <- data,
          sm <- PUSH;
        end;
        when CMD_STACK:
        begin
          match cmd with
          begin
            -- STACK <=> STACK
            when DUP:
            begin
              t0 <- t0',
              sm <- PUSH;
            end;
          ..
          end;
        end;
      end;
      -- Finalizing of stack operations
      match sm with
      begin
        when PUSH:
        begin
          t1 <- t0',
          t2 <- t1',
          tn.[tn_next]<-t2';
          tn_p <- (tn_p + 1) land SS_SIZE_MASK,
          tn_next <- (tn_next + 1) land SS_SIZE_MASK;
          if CHECK_STACK = true then
```

```
70          begin
71            tn_c <- tn_c + 1;
72            if tn_c >= SS_SIZE then raise STACK_OVERFLOW;
73          end;
74        end;
75    .....
76    end;
77  end;
```

The typical execution time for one FORTH word instruction takes less than 10 clock cycles. Compared with a non-optimized von-Neumann architecture processor, requiring 6 clock cycles for the execution of one low-level machine instruction, the high-level instruction execution outperforms classical microprocessors significantly. The bytecode loop profits from highly parallel data-path operations (e.g., lines 28-36) executed in one clock cycle. Stack operations are performed in two steps: A preparation (lines 42-55) and a finalization (lines 60-74). The main command decoder is partitioned into command classes (Literal, Stack, Control, Morphing, and so on) with sub-decoders.

# VII Conclusions

The ConPro programming language bases on a concurrent communicating sequential multi-process model with inter-process-communication and guarded atomic actions, well suited to implement parallel control and data processing systems. Algorithms can be reused from traditional sequential programming. Common algorithms can be programmed and implemented directly in hardware circuits. The ConPro synthesis tool is capable to implement complex algorithms, like communication protocols or high-level virtualizing processors, requiring parallelism and introducing concurrency on control-path level, efficiently in hardware (below and beyond 1 Million gate complexity). Additionaly, there is a software backend generating software from the same Conpro program with (nearly) the same functional behaviour. A rich set of synchronisation objects are provided, too. Hardware blocks and external components are accessed using a method-based and object-orientated programming model (External Module Interface).

# IX Appendix: ConPro Programming Language

## Data Types

Predefined ordinal data types DT are summarized in **Tab. 10**.

*Tab. 10 Data types DT*

| Statement | Type DT | Description |
|---|---|---|
| int[N] | INT | Signed integer with data with of N bits |
| logic<br>logic[N] | LOGIC | Unsigned integer and logic vector (value set {0, 1, Z, H, L}) with data width of N bits. |
| char | CHAR | Character |
| bool | BOOL | Boolean type (value set {true, false}. |
| value | VALUE | Untyped value (integer, logic, char, bool) with data type and width assigned at compile time (only used in constants) |

## Values

Values used in expressions with data content of registers, variables, and signals are associated with different ordinal data types *DT* depending on the value format, summarized in **Tab. 11**.

*Tab. 11 Values and data types DT*

| Value | Type | Description |
|---|---|---|
| -2, -1, 0, 1 ,2 ,3 ,4, ... | INT | Signed integer (decimal format) |
| 0,1,2,3,4,... | INT, LOGIC | Unsigned integer (decimal format) |
| 0x1,0x2,... | INT, LOGIC | Unsigned integer or logic (hexadecimal format: 0-9, A-F, a-f) |
| 0b110,0b0101,... | INT, LOGIC | Unsigned integer or unsigned logic value (binary format: 0,1) |
| 0l111,0lZZZ | LOGIC | Logic value (logic multi-value format; 0,1,L,H,Z) |
| 'a','A',... | CHAR | Character |
| "abc" | STRING | String (character array) |
| true, false | BOOL | Boolean value |
| nanosec, microsec, millisec, sec | UNIT | Time unit (used with integer values) |
| hz, kilohz, megahz, gigahz | UNIT | Frequency unit (used with integer values) |

## EXPRESSIONS AND ASSIGNMENTS

Expressions are used in assignments, branches, function applications, and loops. There are arithmetic, relational, and boolean/logical operations, as shown in **Tab. 12**.

*Tab. 12 Arithmetic, relational, and boolean/logical (bitwise) operators with applicable data types*

| Operator | Type | Description |
|---|---|---|
| `+,-,*,/` | `INT, LOGIC, CHAR` | Addition,Subtraction (Negation), Multiplication, Division. (CHAR: ASCII code) |
| `< <= > >= = <>` | `INT, LOGIC, CHAR` | Lower than, lower equal than, greater than, greate equal than, equal, not equal. |
| `and, or , xor, not` | `BOOL` | Boolean operators |
| `land, lor , lx-or, lnot` | `INT, LOGIC, CHAR` | Bitwise logical operators |
| `@`<br>`a @ b @ c ..` | `LOGIC` | Bitvector concatenation operator |
| `~`<br>`V~B` | `INT` | $\log_B(V)$ (only in constant definitions allowed) |
| `lsl, lsr`<br>*obj* `lsx` *n* | `INT, LOGIC` | Static and dynamic (non-constant shift parameter *n*) shift operations. Left hand operator must be an object (register, variable, signal). |
| `to_logic`<br>`to_int`<br>`to_char`<br>`to_bool` | `INT, BOOL, LOGIC, CHAR` | Type conversion (only applicable to single objects) |

All operands of an expression must be of the same type. Explicit type conversion can be used only to convert the data type of native objects (register, variable, signal).

Assignments of expression values to data objects are shown in **Def. 17**.

*Def. 17 Assignment of a value (calculated from an expression) to a data object (register, variable, signal, queue, channel).*

```
LHS <- RHS;
x <- expr;
```

Function application (only not recursive) is provided by the function name and an argument list, which can be empty (**Def. 18**). Arguments containing expressions are evaluated before function application. Function applications can be embedded in expressions.

*Def. 18 Function application (f) and procedure execution (p) with arguments, w/o arguments*

```
dst <- f(arg1,arg2,...);
dst <- f();
expr(f(...))
p(arg1,arg2,...);
p();
```

## Data Objects

Sequential data processing requires storage objects (memory). There are registers and variables for data storage. A register is a single memory object mappable to CREW access behaviour, whereas a variable is bounded to a memory block (RAM) with EREW access behaviour. Additional there is a signal object without any storage required for inter-connect of hardware components. Data objects are assigned to a specific data type.

There are data objects with local and global visibility (scope). Local objects can only be accessed by one process, whereas global objects can be accessed by multiple processes concurrently. Concurrent access of those global objects are resolved and serialized with a atomic guarded actions and a scheduler for each object.

Storage objects of structure type (concerning only registers and signals) are split into an independent set of storage objects with each object associated with a structure element. Data object definitions are summarized in **Tab. 13**.

*Tab. 13 Data object definitions (TYPE: structure type)*

| Statement | Data Type DT | Description |
|---|---|---|
| `reg R: DT;`<br>`reg R: DT[N];`<br>`reg R,S,...:DT;` | INT, LOGIC, CHAR, BOOL, TYPE | Creates a register storage object R of data type DT and optional data width N (bits). |
| `block B;`<br>`var V: DT in B;`<br>`var V: DT[N] ...` | INT, LOGIC, CHAR, BOOL, TYPE | Creates a variable storage object V of data type DT and optional data width N (bits). Variables are bound in a RAM block, which must be specified explicitly. RAM data cell width and number of cells are determined at compile time. |
| `sig S: DT;`<br>`sig S: DT[N];`<br>`sig S,T,...:DT;` | INT, LOGIC, CHAR, BOOL, TYPE | Creates a inter-connect signal object S of data type DT and optional data width N (bits). |
| `const C:`<br>`  DT := V;`<br>`const C:`<br>`  DT[N] := V;`<br>`const NC:`<br>`  value := V;` | INT, LOGIC, CHAR, BOOL, VALUE | Creates a constant object C with value V of data type DT (or generic VALUE) and optional data width N (bits). |

## IPC Data Objects

There are predefined data objects implementing synchronized inter-process communication (see tables **Tab. 14** and **Tab. 15**): queue and channels. Though queues and channels are abstract objects they can be used directly in expressions (RHS) and assignment statements (LHS). Queues have a buffer storage depth (size) which can be specified with the parameter `depth`. Channels have always a depth of one and can be buffered or unbuffered only implementing a handshake.

Queues or channels of structure type are split into a set of coupled queues or channels with each object associated with one structure element.

*Tab. 14 IPC data object definitions (TYPE: structure type)*

| Statement | Data Type DT | Description |
|---|---|---|
| `queue Q: DT;`<br>`queue Q: DT[N];`<br>`queue Q,R,..: DT with P=V and ..;` | `int, logic, char, bool, TYPE` | Creates a queue object Q of data type DT with optional parameter settings. |
| `channel Q: DT;`<br>`channel Q: DT[N];`<br>`channel Q,R,..: DT with P=V and ..;` | `int, logic, char, bool, TYPE` | Creates a channel object Q of data type DT with optional parameter settings. |

*Tab. 15 IPC data object parameters*

| Parameter | Value | Description |
|---|---|---|
| `depth` | `1..256 (∞)` | Size of queue buffer. Channel depth is always 1. |

## Abstract Objects

Abstract objects are modified only by a set of defined methods. There are builtin objects (queue, channel) and user abstract objects, with behavioural implementation and method interfaces defined by the EMI ADTO programming language, as shown in **Tab. 16**.

*Tab. 16 Abstract object definitions*

| Statement | Type | Description |
|---|---|---|
| `object O: AT;`<br>`object O: AT with P=V and P=V ...;` | `AT: mutex, semaphore, event, timer, ...` | Creates an abstract object O of type AO with optional parameter list (parameter P and value V). |

## Arrays

Arrays can be created with registers, variables, and signals with a specific array element data type. A register array has still CREW access behaviour for each cell, though dynamic element selectors require addressable array selectors (one for each accessing process).

Additionally arrays can be created with structure and abstract object types, as shown in **Tab. 17**.

*Tab. 17 Array definitions (TYPE: structure type)*

| Statement | Data Type DT | Description |
|---|---|---|
| `array A: reg[I] of DT;`<br>`array A: reg[I,J,K] of DT;` | int, logic, char, bool, TYPE | Creates a register storage array A with I elements of data type DT and optional data width N (bits). Multi-dimensional arrays can be created by extending the size parameter list I,J,K. |
| `array A: var[I] of DT;`<br>`block B;`<br>`array A: var[I] of DT in B;` | int, logic, char, bool, TYPE | Creates a variable storage array A with I elements of data type DT and optional data width N (bits) bounded to a RAM block. The assignment of a specific block is optional. |
| `array A: sig[I] of DT;` | int, logic, char, bool, TYPE | Creates a signal array A with I elements of data type DT and optional data width N (bits). |
| `array A: object AT[I];`<br>`array A: object AT[I] with P=V and P=V ...;` | AT: queue, channel, ... | Creates an abstract object array A with I elements of object type AT and optional parameter list (parameter P with value V). |
| `A.[i] <- A.[j];`<br>`A.[i,j,k] <- A.[x,y,z];` | AT | Access of array elements on left-hand (write) and right-hand (read) side of an assignment and in expressions by using the dot bracket selector. |

## STRUCTURES

There are data and component structure types (records). Data structures are used to bind single data elements semantically coupled to a named type. Component structures define a hardware component interface (port signals). Data structure types can be applied to all data object definitions including queues and channels, as shown in **Tab. 18**.

*Tab. 18 Stucture type definitions and definition of objects of structure type.*

| Statement | Data Type DT | Description |
|---|---|---|
| `type T: {`<br>`e1 : DT;`<br>`e2 : DT;`<br>`.. };` | int, logic, char, bool | Defines a data structure type T with elements `e1`, `e2`, .. of specified data types. Registers, variables, and signal objects of this type can be instantiated. |
| `type T: {`<br>`e1 : N1;`<br>`e2 : N2;`<br>`e3 : N3 to N4;`<br>`e4 : N5 downto N6;`<br>`.. };` | logic | Defines a bit-field data structure type T with elements `e1`, `e2`, .. of specified data width (data type logic) `N1`, `N2` ,.... Registers, variables, and signal objects of this type can be instantiated. |
| `type Tc: {`<br>`e1 : DIR DT;`<br>`e2 : DIR DT;`<br>`.. };`<br>`DIR = {input, output, inout}` | int, logic, char, bool | Defines a component structure interface type `Tc` with elements `e1`, `e2`, .. of specified data types and data/signal flow directions. |
| `reg R: T;`<br>`var V: T;`<br>`...` | TYPE | Defines a data object (register `R`, variable `V`, signal `S`) of the user defined structure type T. |
| `component C: Tc;` | TYPEc | Instantiates a component object with an interface structure type `Tc` . |
| `reg R: T;`<br>`R.e1 <- ..`<br>`X <- R.e2 ..` | TYPE | Access of structure elements and bit fields by using the dot selector. |

## ENUMERATION

*Tab. 19 Enumeration type definitions*

| Statement | Object Type | Description |
|---|---|---|
| **`type`** `e : {`<br>`S1;`<br>`S2;`<br>`..`<br>`};`<br><br>`type e : {..`<br>`} with code=C;`<br><br>`C={one,bin,gray, NAME}` | REGISTER, SIGNAL, VARIABLE | Definition of symbolic enumeration list defining named constants.<br>The value of an enumeration element is calculated at compile time. The first enumeration element has index 1, the seconde 2, and so forth. The final coding style can be set with parameter `code`. |

## Processes

A process is the main execution unit. Entire processes are operating independently and concurrently, but process statements are executed sequentially. A process has a local process space consisting of data and some abstract objects. Inter-process communication and synchronization is performed by using global objects with guarded atomic access. Concurrent access is serialized by a scheduler.

*Tab. 20 Process definition and process control.*

| Statement | Description |
|---|---|
| `process P:`<br>`begin`<br>  *`definitions`*<br>  *`statements`*<br>`end;`<br>`process P: ...`<br>`end with P=V and ..;` | Definition of a process with object definitions (optional) and a sequence of statements. Processes can be parameterized by appling a parameter list (paramter P and value V).<br>Except the main process each process must be started explicitly. |
| `array P: process[N] of`<br>`begin`<br>  *`definitions`*<br>  *`statements`*<br>`end;`<br>`# ≡` *`process number`* | Definition of a process array with object definitions (optional) and a sequence of statements. Processes can be parameterized by appling a parameter list (paramter P and value V).<br>N is the size of array (number of processes to be created). |
| `P.call();`<br>`P.start();`<br>`P.stop();` | Process control statements (process methods).<br>Process calling is a synchronous operation. If a process P1 calls a process P2, the process P1 is blocked untill process P2 has finished his work (by reaching the end state).<br>Process starting and stopping does not block the executing process. It is an asynchronous operation. |

## Functions

Functions are implemented by using processes with additional parameters (global registers) intialized with a value at function application time. The application of a function within an expression or the procedure execution passes (optional) arguments to the parameters and starts the process associated with the function. The calling process is blocked untill the function process finishes (by reaching the end state). A return value is passed back to the calling process (in case of a function).

*Tab. 21 Function definition and function application*

| Statement | Description |
|---|---|
| `function f(p1:DT,..)`<br>`  return (pn:DT):`<br>`begin`<br>`  definitions`<br>`  statements`<br>`end;`<br>`..`<br>`end with P=V and .. ;`<br>`..`<br>`end with inline;` | Definition of a function with (optional) formal parameters `p1`, `p2`,.. and a return parameter `pn`. There is no return statement. The value of teh return parameter must be modfied within the function body.<br>Function and procedures can be parameterized. The inline parameter replaces each function application or procedure execution with the respective statement sequence. Local data objects are shared. |
| `function p(p1:DT,..):`<br>`begin`<br>`  definitions`<br>`  statements`<br>`end;` | Definition of a procedure with (optional) formal parameters `p1`, `p2`,.. but without a return parameter. There is no return statement. |
| `P(v1,v2,...);`<br>`P();`<br>`X <- f(v1,v2,..);`<br>`X <- f();` | Function and procedure application with and w/o arguments. |

## Blocks

Process or function statements can be grouped with a block statement. A block can be parameterized, for example with different scheduling, allocation, or optimization behaviour.

*Tab. 22 Statement blocks*

| Statement | Description |
|---|---|
| `begin`<br>`  stmt1;`<br>`  stmt2;`<br>`  ..`<br>`end;` | Sequetntial statement composition. |
| `begin`<br>`  stmt1;`<br>`  stmt2;`<br>`  ..`<br>`end with P=V and ..;` | Sequetntial statement composition with additional behaviour or synthesis parameterization (paremeter P with value V). |
| `begin`<br>`  stmt1;`<br>`  stmt2;`<br>`  ..`<br>`end with bind[=true];`<br>⇔<br>`  stmt1,`<br>`  stmt2,...;` | Parallel data statement composition. Equal to comma separated list of (data assignment only) statements. Only one data statement may perform a guarded access of a global object (read of global registers is not guarded). |

## BRANCHES

There are different branch statements available. They pass the program flow to an alternative statement or a block of statements depending on values, as shown in **Tab. 23**. Branches can appear on module top-level and within block statements (processes, functions...).

*Tab. 23 Branch statements*

| Statement | Kind | Description |
|---|---|---|
| **if** *expr* **then** *statement1*;<br><br>**if** *expr* **then** *statement1*<br>**else** *statement0*; | Boolean Branch | Depending on the result of the boolean expression *expr* a branch occurs either to *statment1* (*expr*=true) or optional to the *statement0* (expr=false). If there is more then one statement, a block is required. |
| **match** *expr* **with**<br>**begin**<br>**when** *v1*: *stmt1*;<br>**when** v1,v2,..:<br>stmt123;<br>**when** v1 to v2:<br>stmtv1tov2;<br>**when** v1 downto v2:<br>stmtv2tov1;<br>**others**: *stmte*;<br>**end**; | Multi-value Matching Branch | Different constant values are matched with the result of the expression *expr* and the respective statements are selected on successfull matching. The default `others` case (matching all other values) is optional.<br>A list of values `v1,v2,..` specifies different alternatives matching the same case. |
| **exception** e1,..;<br>**try**<br>*statement*;<br>**with**<br>**begin**<br>**when** *e1*: *stmt1*;<br>**when** e1,e2,..:<br>stmt123;<br>**others**: *stmte*;<br>**end**; | Exception Handler Branch | Execptions raised in *statements* are matched and the respective statements are selected on successfull matching. The default handler `else` (matching all other exceptions) is optional. |

## Loops

There are different loop statements available. Each loop repeats the execution of the loop body as long as a boolean condition is satisfied. A counting loop iterates a list of values, specified by a range., as shown in **Tab. 24**.

*Tab. 24 Loop statements*

| Statement | Kind | Description |
|---|---|---|
| **for** i =<br>  a to\|downto b<br>**do begin**<br> *statements*<br>**end**;<br><br>end with unroll;<br><br>end with P=V ..; | Counting Loop | The for-loop executes the loop body *statements* for each element in the iterator list,  a range of values including boundaries. The loop iterator variable *i* holds the current ieration value.<br>Loops can be parameterized. The unroll parameters replicates the loop body (b-a)+1 times and replaces the loop iterator with the current iteration value. |
| **while** *expr* **do**<br>**begin**<br>  *statements*<br>**end**; | Conditional Loop | The while-loop executes the loop body as long as the boolean expression *expr* is true. The test of the boolean expression is performed before each loop iteration. |
| **always do**<br>**begin**<br>  *statements*<br>**end**; | Unconditional Loop | This loop never terminates (except by raising an exception). |
| **wait for** *cycles*;<br>**wait for** *time*;<br>**wait for** *cond*;<br>**wait for** *cond*<br> **with** *statement*;<br>**wait for** *cond*<br> **with** *stmt1*;<br> **else** *stmt0*; | Delay/Blocking Loop | The wait for statement blocks the execution until a time interval is passed or a condition is true. Optionally a signal assignement can be applied (stmt1) as long as the condition is false. An optional default statement (stmt0) is applied at the remaining time. |

## Exceptions

Exceptions are used to leave a control environment, for example a function, loop, or branch. Exceptions are propagated beyond control environments until an exception handler (statement) catches the exception, as shown in **Tab. 25**. Otherwise an uncaught exception fault appears.

An exception raised within a nested control environment (nested branches/loops or function calls) is passed to the next higher environment level until a handler environment is reached. Excption handler envionments can be nested, too. Exception not caught by a particular handler (without default-branch) are reraised.

*Tab. 25* *Exception handler statements*

| Statement | Kind | Description |
|---|---|---|
| **exception** ex1,..; | Type | Defintion of a named exception signal. |
| **try**<br>  *statement;*<br>**with**<br>**begin**<br> **when** *e1*: *stmt1;*<br> **when** e1,e2,..:<br>  stmt123;<br> **others**: *stmte;*<br>**end**; | Exception Handler Branch | Execptions raised in *statements* are matched and the respective statements are selected on successfull matching. The default handler `else` (matching all other exceptions) is optional. |
| **raise** ex; | Raising | Raises an exception signal. |

## IPC Objects

Inter-process communication takes place by using global objects. Though global data storage objects are guarded by a mutex scheduler, they are nou suitable for process synchronization. There abstract IPC objects available providing synchronization of different kind, summarized in **Tab. 26**.

*Tab. 26* *Synchronization objects and their methods*

| Object | Methods | Description |
|---|---|---|
| **queue**<br><br>open Core;<br>queue q: DT<br>?with depth=N<br>and scheduler<br>="fifo"; | *q.write*:<br>q <- expr;<br>*q.read*:<br>x <- q;<br>q.unlock() | A queue is a buffer holding up to N elements with synchronized access in FIFO order. The `read` operation blocks until at least one data element is available. The `write` operation blocks if the queue is full. The `unlock` operation unblocks all blocked processes. The data type can either be a core type or record structure type. |
| **channel**<br><br>open Core;<br>channel c: DT<br>?with<br>model=M;<br>M={buffered,<br>unbuffered} | *c.write*:<br>c <- expr;<br>*c.read*:<br>x <- c;<br>c.unlock() | A channel is a buffer holding one (buffered) or no element with synchronized access. The `read` operation blocks until at least one data element is available or one write operation is pending. The `write` operation is blocked until the buffer is empty or a read operation occurs. The `unlock` operation unblocks all blocked processes. |
| **mutex**<br><br>open Mutex;<br>object o: mutex<br>?with<br>schedule=S;<br>S={fifo} | o.lock()<br>o.unlock()<br>o.init() | A mutex implementes mutual exclusion access to shared resources. A mutex is eithe rlcoked or unlocked. Only one process can own the lock by using the `lock` operation. The `unlock` operation releases the lock. Locking an already locked mutex blocks the proecss. Initialization is required by using the `init` operation. |
| **semaphore**<br><br>open Semaphore;<br>object o:<br>semaphore<br>?with<br>scheduler=S and<br>depth=N and<br>init=V;<br>S={"fifo"} | o.down()<br>o.up()<br>o.unlock()<br>o.init(V) | A semaphpore implements asynchronized counter with a value range [0..depth-1] used in producer-consumer applications. The counter value never becoms negative. The `down` opration decrements the counter. If the counter is already zero, the process is blocked. The `up` operation increments the counter. Initialization is required by using the `init` operation. |
| **event**<br><br>open Event;<br>object o: event<br>?with<br>latch=B; | o.await()<br>o.wakeup()<br>o.init() | An event implements simple signal-based process synchronization. Multiple processes can wait for an event by using the `await` operation. Another process can wakeup those blocked processes at the same time by using the `wakeup` operation. A latched event prevent race conditions if the waiting request arrives after the wakeup operation. Initialization is required by using the `init` operation. |
| **barrier**<br><br>open barrier;<br>object o:<br>barrier; | o.await()<br>o.init() | A barrier implements simple signal-based process synchronization of group of N processes. A group of processes can wait for a barrier event (enter the barrier) by using the `await` operation. If the N-th process enters the barrier all processes are unblocked immediately at the same time. The group size N is determined at compile time. Initialization is required by using the `init` operation. |

| Object | Methods | Description |
|---|---|---|
| **`timer`**<br><br>`open Timer;`<br>`object o: timer`<br>`?with`<br>`mode=M;`<br>`M={0,1}` | `o.await()`<br>`o.time(T)`<br>`o.start()`<br>`o.stop()`<br>`o.init()`<br>`o.sig_action(`<br>`S,L1,L0)` | A timer is a self synchronizing event object. Processes can wait for the timer event by using the `await` operation. After the time interval has elapsed, which must be set by the time operation, the waiting processes are woken up. The timer must be started and can be stopped by the `start` and `stop` operations. In `mode=`0 the timer operates continously, in `mode=`1 only one time. The `sig_action` operation attaches a signal S to the timer returning the actual state (L0: inactive, L1: active). |

## Modules

Commonly there is one main module defined by the main entry source file. A module consists explicitly of a module behavioural description including processes, object definitions, and top-level statements and implicitly of a module interface defining the component port interface, at least the clock and reset port signals, as shown in **Tab. 27**. Additional port components can be added by exporting objects (register, signals) and hardware component interfaces.

Additional compound modules can be defined on structural component level. Each compound module conists of an implementation definition (a main module which must be imported), behavioural components instantiated from this main module, and an inter-connect component connecting all instantiated module components.

*Tab. 27 User defined modules*

| Statement | Description |
|---|---|
| *definitions*<br>*declarations*<br>*processes*<br>*top-level stmts* | Declares and creates a top-level behavioural module *M* (from source file *m*.cp). |
| **module** *MC*<br>begin<br>  *import*<br>  *component*<br>  *connect*<br>  *mapping*<br>end; | Declares and creates a new compound module *MC* from a behavioural module with import, component instantiaition, inter-connect and mapping parts. |
| **import** *M*;<br>**component** *C1*, *C2*,..: *M*; | Import of a behavioural module *M* (top-level main module) and module component instantiations (replicated module components). |
| **type** *ic*: {<br>  port *S*: *dir typ*;<br>  ...<br>};<br>**component** *IC* : *ic* := {<br>  C1.S1, ...<br>};<br>IC.S1 << IC.S2; | Definition of an inter-connect component type and instantiation of an inter-connect component with default signal mapping of module component port signals.<br>Finally inter-connect component signals can be connected with additional mapping statements. |